\documentclass[12pt,a4paper]{article}
\usepackage{jheppub}

\newcommand{\de}{\mathrm{d}}

\newcommand{\I}{\mathrm{i}}

\newcommand{\cF}{\mathcal{F}}
\newcommand{\cP}{\mathcal{P}}

\newcommand{\cM}{\mathcal{M}}

\newcommand{\cE}{\mathcal{E}}

\newcommand{\cR}{\mathcal{R}}

\newcommand{\cO}{\mathcal{O}}

\newcommand{\cA}{\mathcal{A}}

\newcommand{\cV}{\mathcal{V}}

\newcommand{\pa}{\partial}

\newcommand{\nn}{\nonumber}

\newcommand{\IR}{\mathbb{R}}

\newcommand{\IZ}{\mathbb{Z}}

\newcommand{\zetastar}{\zeta^\star}

\def\bea{\begin{eqnarray}}
\def\eea{\end{eqnarray}}
\def\be{\begin{equation}}
\def\ee{\end{equation}}
\def\ba{\begin{align}}
\def\ea{\end{align}}
\def\bse{\begin{subequations}}
\def\ese{\end{subequations}}
\def\Im{\,{\rm Im}\,}

\def\no{\nonumber}

\def\RR{{\mathbb R}}

\def\rr{\rho_d}
\def\gd{g_{D}}

\title{$D^{6} \cR^4$ amplitudes in various dimensions}

\preprint{CERN-PH-TH-2015-025}

\author[a,b,c]{Boris Pioline,} 
 
 \affiliation[a]{CERN PH-TH,
Case C01600, CERN, CH-1211 Geneva 23, Switzerland}

\affiliation[b]{Sorbonne Universit\'es, UPMC Universit\'e Paris 6, UMR 7589, F-75005 Paris, France}
 
\affiliation[c]{ Laboratoire de Physique Th\'eorique et Hautes
Energies, CNRS UMR 7589, \\
Universit\'e Pierre et Marie Curie,
4 place Jussieu, 75252 Paris cedex 05, France}

\emailAdd{boris.pioline@cern.ch}

\abstract{Four-graviton couplings in the low energy effective action of type II string vacua compactified on tori are strongly constrained by supersymmetry and U-duality. While the 
$\cR^4$ and $D^4 \cR^4$ couplings are known exactly in terms of Langlands-Eisenstein series of the U-duality group, the $D^6 \cR^4$ couplings are not nearly as well understood. Exploiting the  coincidence of the U-duality group in $D=6$ with the
T-duality group in $D=5$, we propose an exact formula for the $D^6 \cR^4$ couplings in 
type II string theory compactified on $T^4$, in terms of a genus-two modular integral plus a suitable Eisenstein series. The same modular integral computes the two-loop correction to $D^6 \cR^4$
in 5 dimensions, but here provides the  non-perturbative completion of the known perturbative terms in $D=6$. This proposal hinges on a systematic re-analysis of the weak coupling and large radius of the $D^6 \cR^4$ in all dimensions $D\geq 3$, which fills in some gaps and resolves some inconsistencies in earlier studies. 
 }

\begin{document}
\maketitle

\section{Introduction}

Initiated by the seminal work of Green and Gutperle \cite{Green:1997tv}, the analysis of higher derivative corrections to the low energy effective action of flat type II string vacua with maximal supersymmetry has been an invaluable source of insight into the non-perturbative structure of string theory. In dimension $D=10-d$, the moduli space of scalars in these vacua is locally a symmetric space $G/K$, where $G$ is a split real group of type $E_{d+1}$ and $K$ is its maximal 
subgroup \cite{Cremmer:1978ds}. Globally, there is by now overwhelming evidence that vacua related by the action of an arithmetic discrete subgroup $G(\IZ)$ -- known as the U-duality group -- are physically equivalent, as anticipated in \cite{Hull:1994ys}.  This U-duality group 
unifies the T-duality group $SO(d,d,\IZ)$ associated to the $d$-
dimensional internal torus with the global diffeomorphism group $SL(d+1,\IZ)$ manifest in the 
M-theory description \cite{Witten:1995ex} (see Table \ref{Geigentab} and \cite{Obers:1998fb} for a review).
Requiring that the low-energy effective action is invariant under $G(\IZ)$ puts strong constraints both on the possible perturbative and non-perturbative corrections. 

For the leading terms in the low energy expansion, supersymmetry further 
constrains the possible dependence on the moduli \cite{Pioline:1998mn,Green:1998by,Green:2010wi,Basu:2011he,Bossard:2014lra,Bossard:2014aea}, to the extent that they can sometimes be completely determined, to all orders in the string coupling, in terms of suitable automorphic functions on $G/K$. This approach has led to the complete determination of the four-graviton $\cR^4$ and $D^4 \cR^4$ couplings in the low-energy effective action of type II strings compactified on a $d$-dimensional torus down to any dimension $D\geq 3$ \cite{Green:1997di,Kiritsis:1997em,Pioline:1997pu,Obers:1999um,Basu:2007ru,Green:2010wi,Pioline:2010kb,Green:2010kv,Green:2011vz}. Indeed, the functions of the moduli multiplying
these interactions, denoted conventionally by $\cE^{(d)}_{(0,0)}$ and $\cE^{(d)}_{(1,0)}$,
are identified as suitable Langlands-Eisenstein series for the U-duality group (or residues thereof, depending on the normalization convention).
As required by supersymmetry, these automorphic forms are eigenmodes of the Laplace operator on $G/K$, up to certain harmonic anomalies \cite{Green:2010wi},
 \bea
\label{eisenone}
\left ( \Delta_{E_{d+1}} - \frac{3(d+1)(2-d)}{(8-d)} \right )\,
\cE^{(d)}_{(0,0)} & = & 6\pi\, \delta_{d,2}\ ,
\\
\left ( \Delta_{E_{d+1}} - \frac{5(d+2)(3-d)}{(8-d)} \right )\,
\cE^{(d)}_{(1,0)} & = & 40 \, \zeta(2)\, \delta_{d,3} + 7\, \cE_{(0,0)}\, \delta_{d,4}\ .
\label{eisentwo}
\eea
The anomalous terms on the r.h.s. arise in dimensions where ultraviolet divergences in supergravity set in \cite{Green:2010sp}.
Moreover, the  asymptotic expansion of the Eisenstein series at weak coupling regime reproduces the known perturbative corrections \cite{Green:1999pv,D'Hoker:2005ht,Green:2008uj}, along with an infinite series of non-perturbative contributions coming from Euclidean D-branes wrapping cycles of the internal manifold, preserving the expected number of 
supersymmetries \cite{Pioline:2010kb,Green:2011vz}.  In the limit where the radius  of one circle in $T^d$ becomes infinite, $\cE^{(d)}_{(0,0)}$ and $\cE^{(d)}_{(1,0)}$ relate to their higher-dimensional counterparts $\cE^{(d-1)}_{(0,0)}$, $\cE^{(d-1)}_{(1,0)}$ as required by unitarity, whereas, in the limit where the M-theory torus $T^{d+1}$ decompactifies, they can be matched to perturbative computations in eleven-dimensional supergravity \cite{Green:1997as,Green:1997me,Green:1999pu,Green:2005ba,Green:2006gt,Green:2008bf,Basu:2014hsa}.

\begin{table}
$$\begin{array}{|c|c|c|c|c|c|c|c|}
\hline
D & d & G=E_{d+1} & K & {\rm dim}(G/K) & \cE^{(d)}_{(0,0)} & \cE^{(d)}_{(1,0)} & \cE^{(d)}_{(0,1)} \\ \hline
10& 0 & SL(2) & U(1) & 2 & \frac{3}{4} & \frac{15}{4} & 12 \\
 9 & 1 & \IR^+ \times SL(2) & U(1) & 3& \frac{6}{7} & \frac{30}{7} & \frac{90}{7} \\
 8 & 2 & SL(3)\times SL(2) & SU(2) \times U(1) & 7 & \underline{\bf 0} & \frac{10}{3} & 12 \\
 7 & 3 & SL(5) & SO(5) & 14 & -\frac{12}{5} &  \underline{\bf 0} & \frac{42}{5} \\
 6 & 4 & SO(5,5) & SO(5)\times SO(5) & 25 & {\bf -\frac{15}{2}} & {\bf -\frac{15}{2}}  & \underline{\bf 0} \\
 5 & 5 & E_{6(6)} & USp(8)& 42 & {\bf -18} & -\frac{70}{3} & {\bf -18} \\
 4 & 6 & E_{7(7)} & SU(8)& 70 & -42 & {\bf -60} & {\bf -60} \\
 3 & 7 & E_{8(8)} & SO(16) & 128 & -120 & -180 & -198 \\
 \hline
\end{array}
$$
\caption{U-duality group in type IIB string theory compactified on $T^d$. The last three columns tabulate the `eigenvalues' of the $\cR^4$, $D^4\cR^4$ and $D^6\cR^4$ couplings under the Laplacian on the moduli space $G/K$. The boldface highlights degenerate or zero eigenvalues, which are correlated with the onset of infrared divergences, manifested by an anomalous term on the r.h.s. of the Laplace equation. \label{Geigentab}}
\end{table}

While the next term in the low-energy expansion, $D^6 \cR^4$, is still protected by supersymmetry, the exact determination of the function $\cE^{(d)}_{(0,1)}$ multiplying it has been possible so far only in dimension $D\geq 8$, and in quite an implicit way \cite{Green:2005ba,Basu:2007ck,Green:2010wi,Green:2010sp,Green:2010kv,Green:2014yxa}. Part of the difficulty lies in the fact that it receives rather complicated two-loop and three-loop corrections, which have been computed only recently \cite{D'Hoker:2013eea,D'Hoker:2014gfa,Gomez:2013sla}. A second difficulty is  that, unlike the $\cR^4$ and $D^4 \cR^4$ cases, supersymmetry does not require $\cE^{(d)}_{(0,1)}$ to be an eigenmode of the Laplacian,
rather it must satisfy the Poisson equation \cite{Green:2005ba,Green:2010wi}
\be
\begin{split}
\left( \Delta_{E_{d+1}} -{6(4-d) (d+4)\over 8-d} \right)\, \cE^{(d)}_{(0,1)}
=&  -\left ( \cE^{(d)}_{(0,0)} \right)^2  \\
& -\beta_6 \,  \delta_{d,4} 
-\beta_5\, \cE^{(5)}_{(0,0)}\, \delta_{d,5} - \beta_4\, \cE^{(6)}_{(1,0)}\, \delta_{d,6}
\label{eisenthree}
\end{split}
\end{equation}
where the right-hand side involves the {\it square} of the $\cR^4$ coupling, plus anomalous terms when ultraviolet logarithmic divergences appear in supergravity. We shall later on determine the numerical coefficients $\beta_D$ to be\footnote{The value of $\alpha_6$ (see \eqref{alpha654}) and $\beta_6$ were announced in \cite{D'Hoker:2014gfa}, and are confirmed by the independent analysis of \cite{BossardKleinschmidt2015}.}
\be
\label{beta654}
\beta_6=-40\zeta(3)\ ,\quad 
\beta_5=-\frac{55}{3}\ ,\quad 
\beta_4=-\frac{85}{2\pi}\ .
\ee
Due to the occurrence of the square of the Eisenstein series $\cE^{(d)}_{(0,0)}$ on the r.h.s. of \eqref{eisenthree}, the $D^6 \cR^4$ coupling cannot be a (residue of) Langlands-Eisenstein series, but must instead involve a new kind of automorphic object, which does not seem to have been discussed in the mathematics literature.

The main goal of the present work is to determine the exact $D^6 \cR^4$ couplings 
in dimension $D=6$, by making profit of the fortunate coincidence that the U-duality group
in $D=6$, $SO(5,5,\IZ)$, is also the T-duality group in $D=5$. Namely, we claim that 
the exact $D^6 \cR^4$ couplings in type II string theory compactified on $T^4$ is given by
\be
\label{exactD6R46}
\cE^{(4)}_{(0,1)} = \pi\, \int_{\cF_2} \de\mu_2 \, \Gamma_{5,5,2}\, \varphi(\Omega)
+ \frac{8}{189} \hat E^{SO(5,5)}_{[00001],4}\ ,
\ee
where the first term involves an integral over the fundamental domain of the Siegel upper-half plane of degree 2  of the product of $\Gamma_{5,5,2}$, the genus 2 Siegel-Narain partition function of the unique even-self dual lattice of signature $(5,5)$, times  $\varphi(\Omega)$, the Kawazumi-Zhang invariant of genus 2 Riemann surfaces \cite{Kawazumi,zbMATH05661751}. The first term is nothing else but the two-loop contribution to the $D^6 \cR^4$
couplings in $D=5$ \cite{D'Hoker:2013eea,D'Hoker:2014gfa}. The second term 
$\hat E^{SO(5,5)}_{[00001],4}$ is an ordinary Langlands-Eisenstein series in the spinor representation of $SO(5,5)$. As we shall explain, the Ansatz \eqref{exactD6R46} satisfies 
 the Poisson equation \eqref{eisenthree} by construction  and 
reproduces the expected tree-level, one-loop, two-loop and three-loop contributions. It predicts  the Euclidean D-brane instanton contributions in principle, although we shall not attempt to extract them in this work.  The exact $D^6 \cR^4$ couplings in dimension $D=7$ can be obtained by  degenerating $SO(5,5)$ into $SL(5)$.

While the weak coupling expansion of the Eisenstein series in \eqref{exactD6R46} can be obtained straightforwardly from Langlands' constant term formula, the analogous expansion of the genus 2 modular integral in \eqref{exactD6R46} is  challenging, as it depends on the asymptotics 
of the Kawazumi-Zhang invariant, and will be considered 
elsewhere \cite{PiolineRusso-to-appear}. 
Our strategy in this paper will be instead to obtain it  from the large radius expansion of the two-loop contribution to the $D^6 \cR^4$ couplings in $D=5$, which follows from general constraints on the circle decompactification limit of $D^6 \cR^4$ couplings \cite{Green:2010kv,Green:2011vz}. For this purpose,  we shall reanalyze systematically the weak coupling and large radius limits of $\cR^4$, $D^4 \cR^4$
and $D^6 \cR^4$ couplings in all dimensions $D\geq 3$, filling in some gaps and correcting 
various inconsistencies in the literature. We hope that the results in this work can serve as a jumping board to determine the exact $D^6 \cR^4$ couplings in dimension $D<6$ or other exact
couplings in string theory.

The outline of this work is as follows. In Section \ref{sec_revisit} we review the structure of the
$\cR^4$, $D^4\cR^4$ and $D^6 \cR^4$ couplings in string perturbation theory, the differential equations which constrain them, and their behavior under circle decompactification. In particular, we determine the anomalous terms appearing on the r.h.s. of the differential equations for special values of the dimension, and the coefficients of the logarithmic terms which appear in the weak coupling and large radius limit. In Section \ref{sec_d6r46full}, we show that the proposal \eqref{exactD6R46}
for the exact $D^6\cR^4$ amplitude in $D=6$ passes all available consistency checks, including
differential equation, weak coupling and large radius expansion. In Appendix  \ref{app_Eis}, we collect definitions and useful properties of Langlands-Eisenstein series for $SL(d)$, $SO(d,d)$ and exceptional groups.  In Appendix \ref{sec_alldim}, we provide explicit weak coupling and large 
radius expansions of $\cR^4, D^4\cR^4$ and $D^6 \cR^4$ couplings in all dimensions $D\geq 3$.
The bootstrap computation fixing the anomalous coefficients and the constant terms of the relevant Langlands-Eisenstein series and $D^6\cR^4$ couplings can be found in 
Mathematica worksheets submitted to arXiv along with this article.

\section{Revisiting the $D^{2p}\cR^4$ couplings in various dimensions \label{sec_revisit}}

In this section, we perform a systematic re-analysis of the perturbative expansion and large radius limit  of  the $\cR^4$, $D^4 \cR^4$ and $D^6 \cR^4$ couplings in all dimensions $D\geq 3$, closing some gaps in the literature (a brief review was included in \cite{D'Hoker:2014gfa}, but was restricted to $D\geq 6$). Following \cite{Green:1999pv}, we denote by $\cE^{(d)}_{(m,n)}$ with 
$(m,n)=(0,0),(1,0),(0,1)$ the coefficients multiplying $\cR^4$, $D^4 \cR^4$ and $D^6 \cR^4$ in the  local part of the 1-PI  action in Einstein frame. The notation refers to the fact  that these interactions correspond to 
term proportional to $(s^2+t^2+u^2)^m (s^3+t^3+u^3)^n t_8 t_8 \cR^4$ in the low energy 
expansion of the four-graviton scattering amplitude, where $t_8 t_8 \cR^4$ is the standard
contraction of four Riemann tensors which arises at tree level \cite{Green:1981ya,Gross:1986iv}.
The 1-PI action also contains non-local terms
due to the exchange of massless states, which can mix with the local part for particular values of the space-time dimension. Since the Einstein frame metric is invariant under U-duality, the couplings 
$\cE^{(d)}_{(m,n)}$ must be automorphic functions of the moduli in $G/K$. We shall focus on
their expansion at weak coupling and at large radius. The M-theory limit is also interesting but unnecessary for our purposes.

\subsection{Weak coupling limit}

In the weak coupling limit, the scalar moduli space decomposes into
\be
G/K = \IR^+ \times \frac{SO(d,d)}{SO(d)\times SO(d)}\times \IR^{ {\rm dim}(G/K) - d^2-1} \ ,
\ee
where the first factor corresponds to the string coupling $g_D$, the second to the
constant metric and two-form $\rho_d=G+B$ on the torus $T^d$, and the last factor to the Ramond potentials 
when $D>4$, as well as dual of the Neveu-Schwarz fields when $D\geq 4$. The D-dimensional string coupling $g_D$ is related to ten-dimensional type IIB string coupling $g_s$ via $1/g_D^2 = V_d/(l_s^d g_s^2)$, where $V_d$ is the volume of the torus, and is invariant under T-duality.

In string perturbation theory, the four-graviton scattering amplitude is an infinite sum of genus $h$ amplitudes, weighted by $g_D^{2h-2}$, invariant under T-duality at each loop order. After expanding at low energy, and transforming from the string frame to the Einstein frame, the weak coupling expansions of the four-graviton couplings are of the form 
\bea
\label{Edwc1}
\cE^{(d)}_{(0,0)} &=&  \cE_{(0,0)}^{(d),{\rm non. an.}}(\gd,\rr) + \gd^{\frac{2d-4}{d-8}} 
\sum _{h=0} ^\infty \gd^{-2+2h} \, \cE_{(0,0)} ^{(d,h)} (\rr)+ \cO (e^{-2\pi/\gd})
\\
\label{Edwc2}
\cE^{(d)}_{(1,0)} &=&  \cE_{(1,0)}^{(d),{\rm non. an.}}(\gd,\rr) + \gd^{\frac{2d+4}{d-8}} 
\sum _{h=0} ^\infty \gd^{-2+2h} \, \cE_{(1,0)} ^{(d,h)} (\rr)+ \cO (e^{-2\pi/\gd})
\\
\label{Edwc3}
\cE^{(d)}_{(0,1)} &=&  \cE_{(0,1)}^{(d),{\rm non. an.}}(\gd,\rr) + \gd^{\frac{2d+8}{d-8}} 
\sum _{h=0} ^\infty \gd^{-2+2h} \, \cE_{(0,1)} ^{(d,h)} (\rr)+ \cO (e^{-2\pi/\gd})
\eea
where $\cE_{(m,n)} ^{(d,h)} (\rr)$ denotes the $h$-loop contribution and the last term denotes non-perturbative D-brane instanton corrections  (along with  NS-brane instantons when $D\leq 4$) . The first term 
$\cE_{(m,n)}^{(d),{\rm non. an.}}$ is a non-analytic term in the string coupling $g_D$, which 
may arise in the process of transforming from string frame to Einstein frame in the particular 
dimensions where the non-local and local part of the 1-PI effective action mix \cite{Green:2010sp}.
Each of these terms are separately invariant under T-duality. From a mathematical viewpoint,
the expansions \eqref{Edwc1}--\eqref{Edwc3} correspond to the constant term of the automorphic forms $\cE^{(d)}_{(m,n)}$ with respect to the maximal parabolic subgroup $P_1$, obtained by deleting the simple root $\alpha_1$ associated to the `string multiplet'. 

\subsubsection{Perturbative contributions}

As far as the perturbative contributions are concerned, it is by now firmly established that they vanish
but for  the first few loop orders, namely
\be
\cE _{(0,0)} ^{(d,h>1)} = \cE _{(1,0)} ^{(d,h>2)} = \cE _{(0,1)} ^{(d,h>3)} = 0 \ .
\ee
The tree-level contributions are known since \cite{Green:1981ya,Gross:1986iv}, and are independent of the torus moduli,
\be
\label{treecoeffs}
\cE_{(0,0)} ^{(d,0)}  = 2 \zeta(3) \ ,\quad
\cE_{(1,0)} ^{(d,0)}  =  \zeta(5) \ ,\quad
\cE_{(0,1)} ^{(d,0)}  = \frac{2}{3} \zeta(3)^2\ .
\ee
The one-loop contributions are given by modular integrals over the fundamental domain 
$\cF_1$ of the Poincar\'e upper-half plane,
\bea
\label{rfournew}
\cE_{(0,0)} ^{(d,1)} (\rr) & =& \pi\, \int_{\cF_1} \de \mu_1\, \Gamma_{d,d,1} (\rr;\tau)  
\\
\label{dfourrfournew}
\cE_{(1,0)} ^{(d,1)} (\rr) & =& 
2\pi \, \int_{\cF_1} \de \mu_1 \, \Gamma_{d,d,1}(\rr;\tau) \, E^\star(2,\tau)
\\
\label{dsixrfournew}
\cE_{(0,1)} ^{(d,1)} (\rr)  &=& 
\frac{\pi}{3}\, \int_{\cF_1} \de \mu_1 \, \Gamma_{d,d,1}(\rr;\tau) \, \left( 5
E^\star(3,\tau) +\zeta(3) \right)
\eea
where $\Gamma_{d,d,1}$ is the partition function of the Narain lattice at genus 1, and 
 \be
\label{eisndef}
E^\star(s,\tau)=\frac12 \pi ^{-s} \Gamma (s) \zeta (2s)\sum_{(c,d)=1} \frac{(\Im \tau)^s}{|c\tau+d|^{2s}}
\ee
is the non-holomorphic Eisenstein series of $SL(2,\IZ)$, in the normalization of 
\cite{Angelantonj:2011br}.   In defining these divergent integrals we use the renormalization prescription of \cite{MR656029,Angelantonj:2011br}, and normalize the integration measure 
$\de\mu_h$ as in \cite{D'Hoker:2014gfa}. 

At two-loop, the corrections to $\cR^4$ couplings vanish, but the corrections to $D^4\cR^4$ and $D^6\cR^4$ are given by modular integrals over the fundamental domain $\cF_2$ of the Siegel upper-half plane of  degree 2,  which parametrizes genus 2 Riemann 
surfaces, \cite{D'Hoker:2005ht,D'Hoker:2013eea}, 
\bea
\label{dfourrfour2}
\cE_{(1,0)} ^{(d,2)} (\rr)  &=& \frac{\pi}{2} \, \int_{\cF_2}  d \mu_2 \, \Gamma_{d,d,2} ( \rr; \Omega)
\\
\label{dsixrfour2}
\cE_{(0,1)} ^{(d,2)} (\rr)  
&=& \pi\, \int_{\cF_2} d \mu_2 \, \Gamma_{d,d,2}(\rr; \Omega) \, \varphi(\Omega)
\eea
where $\Gamma_{d,d,2}$ denotes the partition function of the Narain lattice at genus 2, and
$\varphi(\Omega)$ is the Kawazumi-Zhang invariant introduced in \cite{zbMATH05661751,Kawazumi}. 

Finally, at three-loop the corrections to $\cR^4$ and $D^4\cR^4$ couplings vanish, but the
correction to $D^6\cR^4$ is given by a modular integral over the fundamental domain $\cF_3$
of the Siegel upper-half plane of  degree 3,  parametrizing genus 3 Riemann 
surfaces, \cite{Gomez:2013sla,D'Hoker:2014gfa}:
\be
\cE_{(0,1)} ^{(d,3)} (\rr)  =\frac{5}{16}\, \int_{\cF_3}  \de \mu_3 \,
\Gamma_{d,d,3}
\label{dsixgenthree}
\ee
where $\Gamma_{d,d,3}$ denotes the partition function of the Narain lattice at genus 3. The normalization here has been fixed by requiring  for $d=0$  the correct value  
$4\zeta(6)/27$ demanded by
S-duality \cite{Green:2014yxa}.

In all cases but the two-loop $D^6 \cR^4$ amplitude, the modular integrals appearing above can be expressed in terms of residues of Langlands-Eisenstein series for the T-duality 
group $SO(d,d,\IZ)$ \cite{Obers:1999um,Green:2010wi,Angelantonj:2011br,Pioline:2014bra}.
Using the conventions for Eisenstein series spelled out in Appendix \ref{app_Eis}, we have
\bea
\label{E001gen}
\cE_{(0,0)}^{(d,1)} &=&2\pi^{2-\tfrac{d}{2}} \Gamma(\tfrac{d}{2}-1)\, E^{SO(d,d)}_{[10^{d-1}],\tfrac{d}{2}-1}\qquad (d\neq 2)
\\
\label{E101gen}
\cE_{(1,0)}^{(d,1)} &=& \frac{2}{45} \pi^{2-\tfrac{d}{2}} \Gamma(1+\tfrac{d}{2})\, 
E^{SO(d,d)}_{[10^{d-1}],\tfrac{d}{2}+1}\qquad (d\neq 4)
\\
\label{E011gen}
\cE_{(0,1)}^{(d,1)} &=& \frac{\zeta(3)}{3}\, \cE_{(0,0)}^{(1)} +
\frac{4}{567} \pi^{2-\tfrac{d}{2}} \Gamma(\tfrac{d}{2}+2)\, 
E^{SO(d,d)}_{[10^{d-1}],\tfrac{d}{2}+2}  \qquad (d\neq 6)
\\
\label{E102gen}
\cE_{(1,0)}^{(d,2)}& =& \frac23 \left( \hat E^{SO(d,d)}_{[0^{d-1}1],2}+ \hat E^{SO(d,d)}_{[0^{d-2}10],2} \right)
\qquad (d\leq 4)
\\
\label{E013gen}
\cE_{(0,1)}^{(d,3)} &=& \frac2{27} \left( \hat E^{SO(d,d)}_{[0^{d-1}1],3}+ \hat E^{SO(d,d)}_{[0^{d-2}10],3} \right)
\qquad (d\leq 6)
\eea
If the Eisenstein series has a pole at the stated value of the parameter $s$, these equations continue to hold after subtracting the pole, i.e. by replacing $E\to \hat E$.  In the last two equations, one should instead replace  $\hat E\to E$ {\it and} drop the second Eisenstein series when $s$ does {\it not} correspond to a pole, i.e. for $d\geq 5$ and $d\geq 7$, respectively.

\subsubsection{Non-analytic contributions}

As far as the non-analytic terms are concerned, they occur in cases where the non-local
and local parts of the action can mix. In practice, this can happen when the
 eigenvalue of $\cE_{(m,n)}^{(d)}$ vanishes, or when it becomes degenerate with 
 that of a coupling $\cE_{(m',n')}^{(d)}$ with fewer derivatives. Looking at Table
 \ref{Geigentab}, we see that this occurs in dimension $D=8$ for $\cR^4$ terms,  
 $D=7,6$ for $D^4\cR^4$ 
 and $D=6,5,4$ for $D^6\cR^4$, along with $D=8$ due to the presence of $[\cE_{(0,0)}^{(2)}]^2$
 on the r.h.s. of the equation \eqref{eisenthree}. Thus, we expect
\be
\label{Enonan}
\begin{split}
\cE_{(0,0)}^{(d),{\rm non-an.}}  =& \frac{4\pi}{3}\, \log g_8\, \delta_{d,2} \\
\cE_{(1,0)}^{(d),{\rm non-an.}}  =& \frac{16\pi^2}{15}\, \log g_7\, \delta_{d,3} + \cE_{(0,0)}^{(4)}\, \log g_6\, \delta_{d,4}\\
\cE_{(0,1)}^{(d),{\rm non-an.}}  =&  \left( \frac{4\pi^2}{27} \log^2 g_8 
+ \frac{2\pi}{9} \left( \frac{\pi}{2}+ \cE^{(2),{\rm an}}_{(0,0)} \right) \, \log g_8 \right)\, \delta_{d,2} \\
& +  \alpha_6 \, \log g_6\, \delta_{d,4} +\alpha_5\, \cE^{(5)}_{(0,0)} \log g_5 \, \delta_{d,5}+\alpha_4\, \cE^{(6)}_{(1,0)} \log g_4 \, \delta_{d,6}
\end{split}
\ee
The coefficients $\alpha_D$ are unknown at this stage, but we shall determine them later on to be 
\be
\label{alpha654}
\alpha_6= 5\zeta(3)\ ,\quad  \alpha_5 = \frac{20}{9}\ ,\quad \alpha_4=\frac{5}{\pi}\ .
\ee
The numerical coefficients in the first three lines have been fixed from the known exact results, although they could be kept as free parameters and fixed in the same way as the 
coefficients $\alpha_D$. The coefficient $\alpha_6$ was erroneous in \cite{Green:2010sp}, 
which caused an apparent discrepancy with the 3-loop supergravity computation 
of \cite{Bern:2008pv}, but the value $5\zeta(3)$ obtained herein resolves this discrepancy,
as already announced in \cite{D'Hoker:2014gfa}. It would
be interesting to similarly check the coefficients $\alpha_5$ and $\alpha_4$ against supergravity
computations.  In \eqref{Enonan}, we have also omitted possible constant terms, which 
can be absorbed in the definition of $g_D$, or equivalently into a different splitting of the 1PI action into  local and non-local parts.

\subsubsection{Differential equations at fixed loop order}

Given the weak coupling expansions \eqref{Edwc1}--\eqref{Edwc3}, 
it is straightforward to translate the differential equations \eqref{eisenone}--\eqref{eisenthree} into  Laplace or Poisson equations for the 
perturbative contributions $\cE_{(m,n)}^{(d,h)}$. The anomalous terms on the r.h.s. of the resulting equations depend on the a priori unknown coefficients $\alpha_D$ and $\beta_D$ in 
\eqref{eisenthree} and \eqref{Enonan} (as well as the `known' coefficients in \eqref{eisenone},
\eqref{eisentwo} and \eqref{Enonan}, which we could keep as free parameters at this stage).
For convenience, we shall display the result only for the relevant values stated in \eqref{beta654} and 
\eqref{alpha654}, which we will derive later on.

Decomposing the Laplacian $\Delta_{E_{d+1}}$ in terms of the $SO(d,d,\RR)$ subgroup,
\be
\label{deltatoso}
\Delta_{E_{d+1}} = \frac{8-d}{8} \pa_\phi^2 + \frac{d^2-d+4}{4}\ \pa_\phi + \Delta_{SO(d,d)} + \cdots
\ee
and using 
\be
\Delta_{E_{d+1}} ( F\,\log \gd ) = \log \gd\, \Delta_{E_{d+1}}  F
+\left( \frac{d^2-d+4}{4} + \frac{8-d}{4} \gd \pa_{\gd} \right)  F
\ee
we find  the following  differential equations for the perturbative terms
$\cE_{(m,n)} ^{(d,h)}$:
\begin{itemize}
\item
\noindent The perturbative corrections to $\cR^4$ couplings satisfy
\bea
\label{delsor4}
\Delta_{SO(d,d)}  ~ \cE_{(0,0)} ^{(0,0)}  & = &  0
\no \\
\left( \Delta_{SO(d,d)}  + d(d-2)/2 \right)\, \cE_{(0,0)} ^{(d,1)}  & = & 4\pi\, \delta_{d,2} 
\eea
\item The perturbative corrections to $D^4\cR^4$ couplings satisfy 
\bea
\Delta_{SO(d,d)}   ~ \cE_{(1,0)} ^{(0,0)}  & = &  0 
\no \\ 
\left( \Delta_{SO(d,d)}  +(d+2)(d-4)/2 \right)\, \cE_{(1,0)} ^{(d,1)} & = & 12\zeta(3)\, \delta_{d,4}
\no \\
\left( \Delta_{SO(d,d)}  + d(d-3) \right)\, \cE_{(1,0)} ^{(d,2)}  & = & 24\zeta(2)\, \delta_{d,3}
+4 \cE_{(0,0)}^{(d,1)} \delta_{d,4}
\label{delsod4r4}
\eea
\item The perturbative corrections to $D^6 \cR^4$ couplings satisfy
\bea
\label{delsod6r40}
\left( \Delta_{SO(d,d)}  -6 \right)\, \cE_{(0,1)} ^{(d,0)}  &=&  -  \left ( \cE_{(0,0)} ^{(d,0)} \right ) ^2
\nn\\
\label{delsod6r41}
\left( \Delta_{SO(d,d)}  - (d+4)(6-d)/2 \right)\, \cE_{(0,1)} ^{(d,1)} &=&
-2 \cE_{(0,0)} ^{(d,0)}  \, \cE_{(0,0)} ^{(d,1)}  +\frac{2\pi}{3} \zeta(3)\, \delta_{d,2}
+\frac{25}{\pi}\zeta(5)\delta_{d,6}
\nn\\
\label{delsod6r42}
\left( \Delta_{SO(d,d)}  - (d+2)(5-d) \right)\, \cE_{(0,1)} ^{(d,2)}  &=&  - \left ( \cE_{(0,0)} ^{(d,1)} \right ) ^2 -\left(\frac{\pi}{3} \cE_{(0,0)}^{(d,1)} + \frac{7\pi^2}{18} \right) \, \delta_{d,2} \nn \\
&& + \frac{70}{3}\zeta(3) \delta_{d,5} + \frac{20}{\pi}  \cE_{(1,0)}^{(6,1)}  \delta_{d,6} 
\nn\\
\label{delsod6r43}
\left( \Delta_{SO(d,d)}  - 3 d(4-d)/2  \right)\, \cE_{(0,1)} ^{(d,3)}   &=&  20 \zeta(3)\, \delta_{d,4} 
+\frac{25}{3} \cE_{(0,0)} ^{(5,1)}  \delta_{d,5}
+\frac{15}{\pi} \cE_{(1,0)} ^{(6,2)} \delta_{d,6}\nn \\&&
\eea 
\end{itemize}
The `eigenvalues' appearing on the l.h.s. of these equations are tabulated in Table \ref{Ttable}.
Except for the two-loop correction to $D^6 \cR^4$, these equations can all be checked against
the Eisenstein series representation of the corresponding amplitude. The equations satisfied by the two-loop modular integrals \eqref{dfourrfour2}, \eqref{dsixrfour2} will be checked 
elsewhere \cite{PiolineRusso-to-appear}.

\begin{table}
$$
\begin{array}{|c|c|c|cc|ccc|}
\hline
D & d & \cE_{(0,0)} ^{(d,1)} & \cE_{(1,0)} ^{(d,1)} & \cE_{(1,0)} ^{(d,2)} & 
\cE_{(0,1)} ^{(d,1)} & \cE_{(0,1)} ^{(d,2)} & \cE_{(0,1)} ^{(d,3)}\\
\hline
 9 & 1 & \frac{1}{2} & \frac{9}{2} & 2 & \frac{25}{2} & 12 & \frac{9}{2} \\
 8 & 2 & {\bf 0} & 4 & 2 & {\bf 12} & {\bf 12} & 6 \\
 7 & 3 & -\frac{3}{2} & \frac{5}{2} & {\bf 0} & \frac{21}{2} & 10 & \frac{9}{2} \\
 6 & 4 & -4 & {\bf 0} & -4 & 8 & 6 & {\bf 0} \\
 5 & 5 & {\bf -\frac{15}{2}} & -\frac{7}{2} & -10 & \frac{9}{2} & {\bf 0} & {\bf -\frac{15}{2}} \\
 4 & 6 & -12 & {\bf -8} &{\bf  -18} & {\bf 0} & {\bf -8} & {\bf -18} \\
 3 & 7 & -\frac{35}{2} & -\frac{27}{2} & -28 & -\frac{11}{2} & -18 & -\frac{63}{2} \\
 2 & 8 & -24 & -20 & -40 & -12 & -30 & -48 \\
 1 & 9 & -\frac{63}{2} & -\frac{55}{2} & -54 & -\frac{39}{2} & -44 & -\frac{135}{2} \\
 \hline
\end{array}
$$
\caption{
Eigenvalues of the perturbative contributions under the T-duality invariant Laplacian $\Delta_{SO(d,d)}$. The degeneracies between different eigenvalues, or their vanishing, are highlighted in boldface, and correlated with the appearance of anomalous terms on the r.h.s. of the Laplace or Poisson equation. 
\label{Ttable}}
\end{table}

\subsection{Circle decompactification limit}

We now turn to the limit in which the radius of one circle in $T^d$, say $r_d$, becomes very large in units of the $D+1$-dimensional Planck scale $l_{D+1}$. This limit is particularly important, as it allows to recursively determine the constant parts of the $D^{2p}\cR^4$ couplings in any dimension from their value in ten-dimensional type IIB theory (or conversely, determine all
of them from their value in $D=3$).  As explained in \cite{Green:2010wi}, in this limit  the coupling $\cE_{(m,n)}^{(d)}$ reduces to its higher-dimensional counterpart $\cE_{(m,n)}^{(d-1)}$ (up to a power of $r_d/l_{D+1}$ determined by dimensional analysis), plus a combination of couplings
$\cE_{(m',n')}^{(d-1)}$ with fewer derivatives, 
\bea
\label{decompR4}
\cE_{(0,0)}^{(d)} &=& \left(\tfrac{r_d}{l_{D+1}}\right)^{\tfrac{6}{8-d}} \left[ \cE_{(0,0)}^{(d-1)} 
+ a_d\, \left(\tfrac{r_d}{l_{D+1}}\right)^{d-3} \right] + \dots 
\\
\label{decompD4R4}
\cE_{(1,0)}^{(d)} &=& \left(\tfrac{r_d}{l_{D+1}}\right)^{\tfrac{10}{8-d}} \left[ \cE_{(1,0)}^{(d-1)} 
+b_d\,  \left(\tfrac{r_d}{l_{D+1}}\right)^{d-5}\, \cE_{(0,0)}^{(d-1)} 
+ c_d\, \left(\tfrac{r_d}{l_{D+1}}\right)^{d+1} \right] + \dots 
\\
\label{decompD6R4}
\cE_{(0,1)}^{(d)} &=& \left(\tfrac{r_d}{l_{D+1}}\right)^{\tfrac{12}{8-d}} \left[ \cE_{(0,1)}^{(d-1)} 
+e_d\, \left(\tfrac{r_d}{l_{D+1}}\right)^{d-7}\, \cE_{(1,0)}^{(d-1)} 
+f_d\, \left(\tfrac{r_d}{l_{D+1}}\right)^{d+3} 
\right. \nn \\   && \left. 
+p_d \, \left(\tfrac{r_d}{l_{D+1}}\right)^{d-3} \cE_{(0,0)}^{(d-1)} 
+q_d\, \left(\tfrac{r_d}{l_{D+1}}\right)^{2d-6}  \right] + 
\dots
\eea
From a mathematical viewpoint,
these expansions  correspond to the constant term of the automorphic forms $\cE^{(d)}_{(m,n)}$ with respect to the maximal parabolic subgroup $P_{d+1}$, obtained by deleting the simple root $\alpha_{d+1}$ associated to the `particle multiplet'. 
From a physics point of view, the additional terms beyond $\cE_{(m,n)}^{(d-1)}$ combine with an infinite series of higher-derivative corrections and a non-local term in dimension $D$ to reproduce the necessary non-local term in dimension $D+1$ due to massless thresholds. In particular, the terms proportional to
$a_d, c_d, f_d$ are the first terms $k=0,2,3$ in an infinite series of local interactions
\be
\cA = r^{d-3} \sum_{k\geq 0} \frac{\pi^k}{k!} \zeta^\star(2k+d-2)\, (r^2 s)^k \cR^4\ ,
\ee
which can be resummed into 
\be
\begin{split}
\cA = & r^{d-3} \sum_{k\geq 0} 
\sum_{m=1}^{\infty} \frac{\pi^{1-\tfrac{d}{2}-k} \Gamma(k+\tfrac{d-2}{2})}{k!\,
m^{2k+d-2}}  (\pi r^2 s)^k \,\cR^4  \\
= & \frac{\pi r^{d-3}}{\sin[\frac{\pi}{2}(d-2)]} 
\sum_{k\geq 0} \sum_{m=1}^{\infty} \frac{(-1)^k\, \pi^{1-\tfrac{d}{2}-k}}{k!\, \Gamma(\tfrac{4-d}{2}-k)\,
m^{2k+d-2}}  (\pi r^2 s)^k \,\cR^4  \\
= & \frac{\pi^{2-\tfrac{d}{2}}}{r \sin[\frac{\pi}{2}(d-2)]\, \Gamma(2-\tfrac{d}{2})}\,
  \sum_{m=1}^{\infty} \left(\frac{m^2}{r^2} -s\right)^{1-\tfrac{d}{2}}\, \cR^4
\end{split}
\ee
where we used $\Gamma(x)\Gamma(1-x)=\pi/\sin\pi x$.  The sum over $m$ in the last line can be interpreted as the sum over massive 
thresholds due to Kaluza-Klein states. The missing term $m=0$ in the sum
is provided by the one-loop massless threshold 
$s^{(2-d)/2} \cR^4$ in the non-local action in dimension $D$. At large 
$r$, we can approximate the sum by an integral, and recover the non-local term $s^{(3-d)/2} \cR^4$
in dimension $D+1$. This fixes, for generic $d$, 
\be
a_d = 4\pi \, \zeta^\star(d-2)\ ,\quad c_d=8\pi \, \zeta^\star(4)\, \zeta^\star(d+2)\ ,\quad 
f_d = \frac{20\pi}{3} \zeta^\star(6)\, \zeta^\star(d+4)\ .
\ee
Similarly, the terms proportional to $b_d$ and $e_d$ are part of an infinite series
of terms  which reproduces the subleading massless threshold in $D+1$ dimensions generated from the product of a tree-level and $\cR^4$ interactions, while the term proportional to $e_d$ enters in an infinite series
which sums up to the massless threshold in $D+1$ dimensions generated from the product of a tree-level and $D^4\cR^4$ interactions. This fixes, for generic $d$,
\be
b_d = 2 \zeta^\star(d-4)\ ,\quad 
e_d = \frac{5}{\pi} \zeta^\star(d-6)\ ,\quad
p_d = \frac{2\pi}{3} \zeta^\star(d-2)\ .\quad
\ee
The term proportional to $q_d$ should similarly be part of an infinite series which sums up to the two-loop supergravity threshold in dimension $D+1$. Its value (as well as the value of $p_d$) is fixed from the differential equation 
\eqref{eisenthree} to be 
\be
q_d = \frac{16\pi^2 [ \zeta^\star(d-2)]^2}{(d+1)(6-d)}\  .
\ee
Using the decompactification limit of the Laplacian, 
\be
\Delta_{E_{d+1}} \to \Delta_{E_{d}} +\frac{8-d}{2(9-d)} (r_d\pa_{r_d})^2 +
\frac{d^2-17d+12}{2(9-d)} r_d\pa_{r_d}
\ee
it is straightforward to check that the differential equations \eqref{eisenone}--\eqref{eisenthree} hold in generic dimension $d$, provided they hold in dimension $d-1$.

For particular values of $d$, the coefficients $a_d\dots q_d$ become singular, at the same time as
powers of $(r_d/l_{D+1})$ become equal in \eqref{decompR4}--\eqref{decompD4R4}. This signals
the presence of logarithmic terms, whose coefficient is a priori unknown. Using 
\be
\Delta_{E_{d+1}}(F \log r_d) 
= (\log r_d)\, \Delta_{E_{d+1}} F  + \left[
 \frac{8-d}{9-d} r_d \pa_{r_d}  +
\frac{d^2-17d+12}{2(9-d)} \right]\, F\ ,
\ee
we see that the anomalous terms on the r.h.s. of  \eqref{eisenone}--\eqref{eisenthree} 
in dimension $d$ are related to the anomalous terms in the same equations in 
dimension $d-1$ and to the coefficients of these logarithms. In Appendix \ref{sec_alldim},
we provide the detailed decompactification limits, including the logarithmic terms,
in any dimension $D\geq 3$.

\subsubsection{Interplay of weak coupling and decompactification limits}

It is also useful to analyze the  limit of each term $\cE_{(m,n)}^{(d,h)}$ in the perturbative 
expansions \eqref{Edwc1}--\eqref{Edwc3} in the limit where the radius of one circle of $T^d$ becomes large in string units.  This allows to relate the $\log \gd$ terms in the weak coupling expansion to the $\log r_{d}/l_{D+1}$ terms in the large radius limit. For this purpose, we need to express the $D$-dimensional coupling $\gd$ and the Planck length in $D+1$ dimensions in terms of the $D+1$-dimensional coupling $g_{D+1}$ and string length $l_s$,
\be
g_D = g_{D+1} (l_s/r_d)^{1/2}\ ,\quad 
l_{D+1} = l_s \, g_{D+1}^{2/(9-d)}\ .
\ee
Clearly, all tree level coefficients are independent of $R\equiv r_d/l_s$. Using
\eqref{decompR4}--\eqref{decompD6R4} and \eqref{Edwc1}--\eqref{Edwc3}, 
one finds that the one-loop corrections
to $\cR^4$, $D^4\cR^4$ and $D^6\cR^4$ couplings behave as 
\be
\label{ER1loop}
\begin{split}
\cE^{(d,1)}_{(0,0)} =& R\, \cE^{(d-1,1)}_{(0,0)} + a_d R^{d-2}\\
&  -4\pi\,\log R\, \delta_{d,2}+4\pi R\, \log R\, \delta_{d,3}
\\
\cE^{(d,1)}_{(1,0)} =& R\, \cE^{(d-1,1)}_{(1,0)} + 
2 \zeta(3)\, b_d R^{d-4}  +
c_d \, R^{d+2} \\
& -4\zeta(3)\, \log R\, \delta_{d,4} + 4\zeta(3)\, R \, \log R\, \delta_{d,5}
\\
\cE^{(d,1)}_{(0,1)} =& R\, \cE^{(d-1,1)}_{(0,1)} + 
\zeta(5)\, e_d R^{d-6} +
f_d \, R^{d+4}  +  2 \zeta(3) \, p_d R^{d-2} \\
& +\left( -\frac{4\pi}{3} \zeta(3) \log R + \frac{\pi}{18}\zeta(3) \right)\,\delta_{d,2}
+ \left(\frac{4\pi}{3} \zeta(3) R \log R-\frac{\pi}{18}\zeta(3) R\right)\, \delta_{d,3} \\
&
-\frac{5}{\pi} \zeta(5) \log R\,\delta_{d,6} + \frac{5\zeta(5)}{\pi} R \log R \, \delta_{d,7}
\end{split}
\ee
where the terms proportional to $a_d, b_d,$ etc are to be omitted  in the dimensions where they are singular, and replaced by the explicit logarithmic terms displayed on the subsequent line. These large radius expansions can be checked from the Eisenstein series representations \eqref{E001gen}--\eqref{E011gen}, or just as well from the modular integral representation 
\eqref{rfournew}--\eqref{dsixrfournew}, using e.g. the techniques in \cite{Angelantonj:2011br}.

Similarly, the two-loop corrections to $D^4\cR^4$ and $D^6\cR^4$ behave as 
\be
\label{ER2loop}
\begin{split}
\cE^{(d,2)}_{(1,0)} =& R^2\, \cE^{(d-1,2)}_{(1,0)} + 
b_d R^{d-3}\, \cE^{(d-1,1)}_{(0,0)} \\
&-\frac{4\pi^2}{3}\log R\, \delta_{d,3}
+\left({ \frac{4\pi^2}{3} R^2}
 - 2R \cE^{(3,1)}_{(0,0)}\right) \log R\, \delta_{d,4}
+2R^2\,\log R\, \cE^{(4,1)}_{(0,0)} \,\delta_{d,5}
\\
\cE^{(d,2)}_{(0,1)} =& R^2\, \cE^{(d-1,2)}_{(0,1)} + 
e_d R^{d-5}\, \cE^{(d-1,1)}_{(1,0)} 
+ p_d R^{d-1}\, \cE^{(d-1,1)}_{(0,0)} +
q_d  \, R^{2d-4}\\
& +\left[ \frac{4\pi^2}{3} (\log R)^2 -\frac{\pi^2}{3}\log R
+\left(-\frac{2\pi}{3} R \log R+
\frac{\pi}{36}R\right)\, \cE^{(1,1)}_{(0,0)} + \frac{37\pi^2}{216} \right] \delta_{d,2}\\
& +\left( \frac{4 \pi^2}{3} (R\log R)^2 + \frac{2\pi}{3} R^2 \log R\, 
( \cE^{(2,1)}_{(0,0)} + \frac13 ) + 
\frac{5\pi^2 R^2}{72} -\frac{\pi R^2}{36}\cE^{(2,1)}_{(0,0)}  \right)\delta_{d,3} \\
&
-\frac{10}{3}\zeta(3)\,\log R\, \delta_{d,5}
+ \left( { \frac{10}{3} R^2 \zeta(3)} - \frac{5}{\pi} R \, \cE^{(5,1)}_{(1,0)} - \frac{4\pi^6}{14175} R^8\right)
\,\log R \, \delta_{d,6}  \\
& +\frac{5}{\pi} R^2 \log R\, \cE^{(6,1)}_{(1,0)}\, \delta_{d,7}
\end{split}
\ee
The large radius expansion of $\cE^{(d,2)}_{(1,0)}$ can be checked using the Eisenstein 
series representation \eqref{E102gen}. The expansions can also be checked directly using
the genus 2 modular integral representations \eqref{dfourrfour2}--\eqref{dsixrfour2}, although
for the latter detailed information about the asymptotics of the Kawazumi-Zhang invariant is 
required \cite{PiolineRusso-to-appear}. 

Finally, the three-loop $D^6\cR^4$ correction behaves as 
\be
\label{ER3loop}
\begin{split}
\cE^{(d,3)}_{(0,1)} =& R^3\, \cE^{(d-1,3)}_{(0,1)} + 
e_d R^{d-4}\, \cE^{(d-1,2)}_{(1,0)} -\frac{10}{3}\zeta(3) \log R\, \delta_{d,4}\\
&
+\left( { \frac{10}{3}\zeta(3) R^3} - \frac53 R \,\cE^{(4,1)}_{(0,0)} \right)\, \log R\, \delta_{d,5}
+\left({ \frac53 R^3 \cE^{(5,1)}_{(0,0)} } - \frac5{\pi}R^2 \cE^{(5,2)}_{(1,0)} \right)\, \log R\, \delta_{d,6}\\
& + \frac{5}{\pi} \cE^{(6,2)}_{(1,0)}\, R^3 \log R\, \delta_{d,7}
\end{split}
\ee
as can be checked using the Eisenstein series representation \eqref{dsixgenthree}.

 \subsection{Bootstrap}
 
In the previous subsections, we have assumed specific values for the coefficients of the logarithms appearing in the weak coupling and large radius expansions, and consequently for the anomalous terms in the partial differential equations. We now comment on how these values have been obtained.
For what concerns the large radius, fixed loop order behavior, we have already mentioned that the coefficients of the logarithms could be fixed from the Eisenstein series  representations \eqref{E001gen}--\eqref{E013gen}, except for the $D^6\cR^4$ two-loop correction, which cannot 
be represented as an Eisenstein series. Also, the coefficients of the non-analytic terms in  \eqref{Enonan} are a priori unknown, although they could in principle be fixed by a supergravity computation. By requiring the consistency of the weak coupling and large radius expansions, 
it turns out that all coefficients are fixed uniquely to the values stated above. We refer the interested reader to the Mathematica worksheet {\tt d6r4bootstrap.nb} submitted along with this article on arXiv for details.

\subsection{Dimensional regularization: a puzzle}

While we have followed the bootstrap strategy to fix the coefficients of the logarithms, one could also try to use dimensional regularization to determine these  coefficients. For this purpose, let us 
denote by $\widetilde{\cE}^{(d)}_{(m,n)}$ the $\cR^4$, $D^4\cR^4$ and $D^6\cR^4$ couplings in generic dimension\footnote{Dimensional regularization is tricky to implement in string theory, but the anomalous terms are expected to be determined in supergravity supplemented with suitable 
counterterms.} $D=10-d$, and assume that they satisfy the differential equations \eqref{eisenone}--\eqref{eisenthree} with no anomalous terms (but still with the quadratic source term in \eqref{eisenthree}). Similarly, we assume that the genus $h$ contributions $\widetilde{\cE}^{(d,h)}_{(m,n)}$ are expected to satisfy \eqref{delsor4}--\eqref{delsod6r43} with no anomalous terms. 
We expect $\widetilde{\cE}^{(d)}_{(m,n)}$ to have a pole at 
values of $d$ where the non-local and local actions mix. Defining the finite coupling 
${\cE}^{(d)}_{(m,n)}$ by subtracting the pole,  the differential equation for 
${\cE}^{(d)}_{(m,n)}$ will pick up an anomalous term proportional to the residue at the pole.
For example, the anomalous term on the r.h.s. of the differential equation \eqref{eisenone} 
for the $\cR^4$ coupling in $D=8$ follows if $\widetilde{\cE}^{(d)}_{(0,0)}$
has a simple pole at $d=2$,
\be
\label{ER4pole2}
\widetilde{\cE}^{(d)}_{(0,0)} = -\frac{4\pi}{d-2} + {\cE}^{(2)}_{(0,0)} + \cO(d-2)
\ee
such that 
\be
\left ( \Delta_{E_{d+1}} - \frac{3(d+1)(2-d)}{(8-d)} \right )\, \widetilde{\cE}^{(d)}_{(0,0)} = 0 \ .
\ee
The singularity of $\widetilde{\cE}^{(d)}_{(0,0)}$ at $d=2$ can be further assigned to a simple pole in the one-loop contribution,
\be
\widetilde{\cE}^{(d,1)}_{(0,0)} = -\frac{4\pi}{d-2} + {\cE}^{(2,1)}_{(0,0)} + \cO(d-2)\ ,
\ee
producing the correct anomalous term on the r.h.s. in the second line of \eqref{delsor4}. Taking into
account the poles in the coefficient $a_d$ appearing in \eqref{decompR4} for $d=2$ and $d=3$,
\be
a_d\sim -\frac{4\pi}{d-2} \ ,\quad a_d\sim \frac{4\pi}{d-3}\ ,
\ee
we recover the $\log R$ terms in the second line of \eqref{ER1loop}  with the correct coefficient, as well as the $\log r$ terms in \eqref{ER48R} and \eqref{ER47R}.

Similarly the anomalous terms in  \eqref{eisentwo} seem to imply that 
$\widetilde{\cE}^{(d)}_{(1,0)}$ has poles at $d=3$ and $d=4$,
\be
\label{ED4R4pole3}
\widetilde{\cE}^{(d)}_{(1,0)} = -\frac{4\pi^2}{3(d-3)} + {\cE}^{(3)}_{(1,0)}+\dots 
\ee
\be
\label{ED4R4pole4}
\widetilde{\cE}^{(d)}_{(1,0)} =
-\frac{56}{85} \frac{{\cE}^{(4)}_{(0,0)}}{d-4} +  {\cE}^{(4)}_{(1,0)}+ \dots \ ,\quad
\ee
The pole at $d=3$ originates from simple pole in the two-loop contribution, consistently with the anomalous term in the second line of \eqref{delsod4r4}. It is also consistent with \eqref{decompD4R4} and \eqref{ED4R47R}, upon noting that $b_3=\pi/3$. The pole at $d=4$ is on the other hand puzzling: indeed the anomalous terms in the differential equations \eqref{delsod4r4} seem to require
\be
\widetilde{\cE}^{(d,1)}_{(1,0)} =-\frac{4\zeta(3)}{d-4}+ \dots,\ \quad
\widetilde{\cE}^{(d,2)}_{(1,0)} =-\frac45 \frac{{\cE}^{(4,1)}_{(1,0)}}{d-4} + \dots
\ee
in disagreement with \eqref{ED4R4pole4}. Moreover, using $b_d\sim -2/(d-4)$, the decompactification limit  \eqref{decompD4R4} seems to require  a coefficient $-2$ in 
\eqref{ED4R4pole4}, rather than $-56/85$, while the coefficients of $r_4^{3/2} \log r_4$
and $r_5^{3/2} \log r_4$ in \eqref{ED4R46} differ from the ones predicted by  \eqref{decompD4R4}. 

As for the $D^6\cR^4$ couplings, the differential equations \eqref{eisenthree} seem to 
imply that $\widetilde{\cE}^{(d)}_{(0,1)}$ has poles at $d=2,4,5,6$, 
\be
\begin{split}
\widetilde{\cE}^{(d)}_{(0,1)} =&  
\frac{4\pi^2}{3(d-2)^2} -\frac{2\pi}{3} \frac{ {\cE}^{(2)}_{(0,0)} + {\rm cte} }{d-2} + {\cE}^{(2)}_{(0,1)} 
= 
\frac{10\zeta(3)}{3(d-4)} + {\cE}^{(4)}_{(0,1)}+\dots \\
= &-\frac{55}{78} \frac{{\cE}^{(5)}_{(0,0)}}{d-5} +  {\cE}^{(5)}_{(0,1)}+ \dots   
= -\frac{85}{132\pi} \frac{   {\cE}^{(6)}_{(1,0)} }{(d-6)} +  {\cE}^{(6)}_{(0,1)}+\dots
\end{split}
\ee
however the coefficients of the pole in $d=5$ and $d=6$ are in conflict with the fixed order differential equations and decompactification limits. We leave it as an open problem to resolve these discrepancies, and adopt the results of the bootstrap method, which have been checked thoroughly.

\section{Non-perturbative $D^6 \cR^4$ couplings in $D=6$ and $D=7$ \label{sec_d6r46full}}

In Type II string theory compactified on $T^4$, the T-duality symmetry $SO(4,4,\IZ)$ and the diffeomorphism
group $SL(5,\IZ)$ of the M-theory $T^5$ torus combine into the U-duality group $SO(5,5,\IZ)$. 
The U-duality invariant quadratic form in the fundamental representation is given by
\be
\cM^2 =  \frac{l_M^3}{V_5} (m^I+C^{IK} n_K) g_{IJ} (m^J+C^{JL} n_L) + \frac{V_5}{l_M^3}  n_I g^{IJ} n_J\ .
\ee
where $g_{IJ}$ is the metric on $T^5$, $C^{IJ}=\epsilon^{IJKLM} C_{KLM}$ is the 3-form, $V_5=\sqrt{\det g_{IJ}}$, and $l_M$ is the 11-d Planck scale. This quadratic form provides  (up to an overall factor of $1/l_6^4=V_5/l_M^9$), the square of the tension of a string made out of M2-branes wrapping the 1-cycle $m^I$ and M5-branes wrapping the 4-cycle $\epsilon^{IJKLM} n_M$.
Decomposing $T^5=T^4\times S^1$ and reducing M-theory along the circle  $S^1$ with radius $r_s$, one arrives at type IIA string theory with $r_s=g_s l_s, l_M^3=g_s l_s^3$. In the weak coupling limit $g_s\to 0$, the lattice $\Gamma_{5,5}$ of string charges decomposes into $\Gamma_{4,4}\times \Gamma_{1,1}$, with (for $C^{IJ}=0$)
\be
\cM^2 =  g_6^2 (m^s)^2 + \frac{1}{g_6^2} (n_s)^2  + \frac{l_s^2}{V_4} m^i g_{ij} m^j + 
\frac{V_4}{l_s^2} n_i g^{ij} n^j  
\ ,\quad
\frac{1}{g_6^2} =   \frac{V_5}{r_s^2 l_M^3}=
\frac{V_4}{g_s^2 l_s^4} \ .
\ee
so the effective radius along $\Gamma_{1,1}$ is $1/g_6$. Upon dualizing
the charge $n_i$ into $n^{ijk}$, we see that $(m^i,n^{ijk})$ transform as a spinor of $SO(4,4)$. 

The same arithmetic group $SO(5,5,\IZ)$ also arises as the T-duality group of string theory compactified on $T^5$. In that case, the degeneration $\Gamma_{5,5}\to \Gamma_{4,4}\times \Gamma_{1,1}$ arises upon decompactifiying a circle. In this case, the mass formula for the winding
and momentum states (in string units) decomposes into
\be
\cM^2 = \frac{l_s^2}{r_5^2} (m^5)^2 +  \frac{1}{l_s^2}\,m^a \gamma_{ab} m^b + l_s^2\, n_a \gamma^{ab} n_b
+ \frac{r_5^2}{l_s^2} (n_5)^2\ .
\ee
Now, $(m^a, n_a)$ transform as a vector of $SO(4,4)$. 
The two mass formulae provided we identify $r_5/l_s = 1/g_6$ and $\gamma_{ab}$ with the image of $g_{ij}$ under triality, such that the spinor $(m^i,n^{ijk})$ is mapped to the vector $(m^a,n_a)$. This opens the possibility that  an automorphic form for $SO(5,5,\IZ)$ might represent both a perturbative contribution in type II string theory on $T^5$ at fixed loop order, or a non-perturbatively exact coupling in type II string theory on $T^4$.

The first example of this arises for $\cR^4$ couplings. Indeed, the non-perturbative  $\cR^4$ coupling in $D=6$ is equal to the Eisenstein series 
\be
E^{SO(5,5)}_{[10000],3/2}
= \frac{2\zeta(3)}{g_6^3} + \frac{2}{g_6} \, E^{SO(4,4)}_{[1000],1}
\ee
This is also equal to the one-loop contribution to the $\cR^4$ coupling in $D=5$, which decomposes in the large radius limit as
\be
E^{SO(5,5)}_{[10000],3/2} = 2\zeta(3)\, r_5^3 + 2 r_5\, E^{SO(4,4)}_{[1000],1}
\ee
This is indeed related to the weak coupling limit under $(r_5/l_s) = 1/g_6$ and $SO(4,4)$ triality, since $E^{SO(4,4)}_{[1000],1}$ happens to be invariant under triality \cite{Kiritsis:2000zi}.

Let us apply the same idea to the $D^6 \cR^4$ coupling in $D=6$. The two-loop
contribution to the $D^6 \cR^4$ coupling in $D=5$ satisfies
\be
\Delta_{SO(5,5)} \cE^{(5,2)}_{(0,1)} = - \left[ \cE^{(5,1)}_{(0,0)} \right]^2  + \frac{70}{3} \zeta(3)\ .
\ee
On the other hand, the exact $D^6 \cR^4$ coupling in $D=6$ satisfies
\be
\Delta_{SO(5,5)} \cE^{(4)}_{(0,1)} = - \left[ \cE^{(4)}_{(0,0)} \right]^2 + 40 \zeta(3)
\ee
We can therefore decompose
\be
\cE^{(4)}_{(0,1)} = \cE^{(5,2)}_{(0,1)}  + \cF
\ee
where 
\be
\Delta_{SO(5,5)} \cF = \frac{50}{3} \zeta(3)\ .
\ee
The behavior of $\cE^{(5,2)}_{(0,1)}$  in  the decompactification limit $r_5\to\infty$,
\be
\cE^{(5,2)}_{(0,1)} = 
 \frac23\zeta(3)^2\, r_5^6 
+ \frac{\zeta(3)}{3} r_5^4\, \cE^{(4,1)}_{(0,0)} 
+ r_5^2\, \cE^{(4,2)}_{(0,1)} 
+ \frac{5}{6} \cE^{(4,1)}_{(1,0)} 
-\frac{10}{3}\zeta(3) \log r_5 +\dots
\ee
is interpreted as a weak coupling expansion
\be
\cE^{(5,2)}_{(0,1)} = 
\frac23\zeta(3)^2\,g_6^{-6} 
+\frac{2\zeta(3)}{3} g_6^{-4}  E^{SO(4,4)}_{[1000],1}
+ g_6^{-2}  \check{\cE}^{(4,2)}_{(0,1)} 
+ \frac{2}{27} \hat E^{SO(4,4)}_{[0001],3}+\frac{10}{3}\zeta(3) \log g_6+\dots
\ee
where $\check{\cE}^{(4,2)}_{(0,1)}$ denotes the image of ${\cE}^{(4,2)}_{(0,1)}$ under triality, 
and we have again used the fact that $E^{SO(4,4)}_{[1000],1}$ (but not $\hat E^{SO(4,4)}_{[0001],3}$) was invariant under triality.

Comparing with the desired result \eqref{D6R46},
\be
\begin{split} 
\cE^{(4)}_{(0,1)} = &
\frac23\zeta(3)^2\,g_6^{-6} 
+g_6^{-4} \left[ \frac{2\zeta(3)}{3}  E^{SO(4,4)}_{[1000],1}
+ \frac{8}{189} E^{SO(4,4)}_{[1000],4} \right]
+ g_6^{-2}  {\cE}^{(4,2)}_{(0,1)} \\
&
+ \frac{2}{27} \left[ \hat E^{SO(4,4)}_{[0001],3}+ \hat E^{SO(4,4)}_{[0010],3}\right]
+ 5\zeta(3) \log g_6+\dots
\end{split}
\ee
we see that $\cE^{(5,2)}_{(0,1)}$ correctly reproduces
the tree-level term, part of the one-loop and three-loop terms, and the 
two-loop term, under the condition that 
\be
{\cE}^{(4,2)}_{(0,1)}=\check{\cE}^{(4,2)}_{(0,1)}\ .
\ee
Indeed, one can check that the constant term of ${\cE}^{(4,2)}_{(0,1)}$ with respect to the Borel subgroup of $SO(4,4)$ is invariant under triality. 
The remainder $\cF$ must then produce
\be
\begin{split}
\cF = & \frac{8}{189} g_6^{-4}  E^{SO(4,4)}_{[1000],4}
+\frac{2}{27}  \hat E^{SO(4,4)}_{[0010],3} +\frac{5}{3}\zeta(3) \log g_6+\dots \\
=& \frac{8}{189} r_5^4  E^{SO(4,4)}_{[0001],4}
+\frac{2}{27}  \hat E^{SO(4,4)}_{[0010],3}  - \frac{5}{3}\zeta(3) \log r_5+\dots
\end{split}
\ee
Thus $\cF$ is proportional to the Eisenstein series
\be
\hat E^{SO(5,5)}_{[00001],4} =r_5^4 E^{SO(4,4)}_{[0001],4} + \frac74 
\hat E^{SO(4,4)}_{[0010],3} - \frac{315}{8} \zeta(3) \log r_5
\ee
Altogether, we have therefore obtained the exact $D^6 \cR^4$ coupling in type II string theory compactified on $T^4$,
\be
\label{D6R46full}
\cE^{(4)}_{(0,1)} = \pi\, \int_{\cF_2} \de\mu_2 \, \Gamma_{5,5,2}\, \varphi(\Omega)
+ \frac{8}{189} \hat E^{SO(5,5)}_{[00001],4}\ .
\ee
It would be very interesting to extract the instanton effects in the weak coupling limit, but this will require detailed knowledge of the asymptotics $\varphi(\Omega)$.

To obtain the corresponding result in $D=7$, we should take the limit $r_4/l_7\to \infty$, and extract
the term of order $(r_4/l_7)^3$ in \eqref{ERD6R46}. In this limit, the $SO(5,5,\IZ)$ duality group is
broken to $SL(5,\IZ)$.  the Eisenstein series decomposes into 
\be
\frac{8}{189} \hat E^{SO(5,5)}_{[00001],4}
= \frac{16}{189} \zeta(8)  \left(\frac{r_4}{l_7}\right)^{10} +\frac{5\pi}{378} 
E^{SL(5)}_{[0010],7/2} \left( \frac{r_4}{l_7}\right)^{3} + \frac{5\zeta(3)}{4\pi^2} 
\hat E^{SL(5)}_{[1000],5/2} - \frac52 \zeta(3)\, \log \frac{r_4}{l_7}\ ,
\ee
reproducing part of the terms in \eqref{ERD6R46}.
As for the genus 2 modular integral, viewing $SO(5,5,\IZ)$ as the T-duality group in $D=5$, we have to study the limit when the volume $V_5/l_s^5$ is scaled to infinity, and extract the term of order $(r_4/l_7)^3 = (V_5/l_s^5)^{6/5}$. 
The torus decompactification limit  can be analyzed by applying
the orbit method on the genus 2 Narain partition function $\Gamma_{5,5,2}$, 
following \cite{Pioline:2014bra}. The zero and rank one orbits 
reproduce the $\cO(r_4^5)$ and $\cO(r_4^4)$ terms in \eqref{ERD6R46}, while the rank 2 orbits contributes to the $\cO(r_4^3)$ term. For the rank 2 orbits, the winding numbers can be set 
to zero at the cost of extending the integration domain from $\cF_2$ to the `generalized 
strip' $GL(2,\IZ)\backslash(\cP_2\times [-1/2,1/2[^3)$, where the first factor corresponds to the
imaginary part $\Omega_2$ of the period matrix, valued in the space of positive definite $2\times 2$
matrices $\cP_2$, while the second factor corresponds to the real part $\Omega_1$ whose entries are restricted to the interval $[-1/2,1/2[$.
The integral over $\Omega_1$ then projects the Kawazumi-Zhang invariant to its supergravity limit, given 
by \cite{D'Hoker:2014gfa}
\be
\varphi_L(\Omega_2) = \frac{\pi}{6}
\left( L_1 + L_2 + L_3 - \frac{5 L_1 L_2 L_3}{L_1 L_2+L_2 L_3+L_1 L_3} \right)\ ,
\ee
where $0<L_3\leq L_1\leq L_2$ parametrize $\Omega_2$ in the fundamental domain  of $GL(2,\IZ)\backslash \cP_2$,
\be
\Omega_2 = \begin{pmatrix} L_1+L_3 & L_3 \\ L_3 & L_2 + L_3 \end{pmatrix}\ .
\ee
Thus, we arrive at 
\be
\begin{split}
\cE^{(3)}_{(0,1)} = & \frac{4\pi^2}{3}\, 
\int_{GL(2,\IZ)\backslash \cP_2} \frac{\de^3 \Omega_2}{|\Omega_2|^3}\, 
\sum_{M^i_\alpha} \exp(-\pi \hat g_{ij} M^i_\alpha\, [\Omega_2^{-1}]^{\alpha\beta} M^i_\beta)\,
\varphi_L(\Omega_2) 
+\frac{5\pi}{378} 
E^{SL(5)}_{[0010],7/2} 
\end{split}
\ee
where $\hat g_{ij}$ is the $5\times 5$ unit-determinant positive definite matrix parametrizing the moduli space $SL(5)/SO(5)$ in $D=7$, and the sum runs over $5\times 2$ rank two integer matrices  $M^i_\alpha$ (the dual momenta). This can be rewritten as an integral over $\IR^+\times \cF_1$, using the same change of variables 
as in \cite{Green:1999pu,Green:2005ba}.
It would be interesting to analyze the D-instanton effects, and make contact with the proposal in \cite{Basu:2007ck,Green:2010wi} in $D=8$ by a further circle decompactification. 
In the other direction, it is a challenge to generalize the proposal \eqref{D6R46full} in dimension $D<6$, where the U-duality group becomes exceptional.

\medskip

{\bf \noindent Acknowledgments:} It is a pleasure to thank Guillaume Bossard, Eric d'Hoker, Michael Green, Axel Kleinschmidt and Rodolfo Russo for useful discussions. Special thanks are due to Rodolfo Russo, for collaboration on the study of asymptotics of modular integrals,  to Guillaume Bossard for his help in bringing out the importance of triality in the construction in  Section 3, and to 
Axel Kleinschmidt for advice on  computations of constant terms for Eisenstein series.

 \appendix

\section{A compendium on Langlands-Eisenstein series \label{app_Eis}}

In this section we briefly review the definitions and main properties of Langlands-Eisenstein series associated to maximal parabolic subgroups, following \cite{Green:2010kv,Fleig:2012xa}, and collect useful facts 
about  Eisenstein series for $SL(d)$, $SO(d,d)$, and for the exceptional groups $E_6, E_7, E_8$. 

\subsection{Generalities on Langlands-Eisenstein series}

The general Langlands-Eisenstein series for a finite simply laced Lie group $G$ with arithmetic subgroup $G(\IZ)$ in split real form  is defined  by
\be
\label{defEisLanggen}
\cE^G(\lambda;g) 
= \sum_{\gamma\in B(\IZ)\backslash G(\IZ)} e^{\langle \lambda+\rho, H(\gamma g)\rangle}
\ee
where $\lambda$ is a vector in weight space, $B(\IZ)$ is the intersection of the Borel subgroup $B$ with $G(\IZ)$, $\rho$ is the Weyl vector (the sum of all fundamental weights, or half the sum of all positive roots), and 
$H(g)=(\log t_1,\dots \log t_r)$ is the logarithm of the Abelian part in the Iwasawa decomposition $G=KAN$. The sum is absolutely convergent when the real part of $\lambda$ has sufficiently large positive inner product with all simple roots, and it can be meromorphically continued to all $\lambda$. The meromorphic continuation satisfies the functional equation
\be
\cE^G(\lambda;g)  = M(w,\lambda)\, \cE^G(w\cdot \lambda;g) 
\ee
for any $w$ in the Weyl group $W$ of $G$. Here, $M(w,\lambda)$ is the reflection coefficient
\be
M(w,\lambda) = \prod_{\alpha\in\Delta_+, w\cdot \alpha\in\Delta_-} 
\frac{\zeta^\star( \langle \lambda,\alpha \rangle)}{\zeta^\star(1+ \langle \lambda,\alpha \rangle)}\ ,
\ee
where $\Delta_+$ is the set of positive roots, $\Delta_-=-\Delta_+$, and $\zeta^\star(s)=\pi^{-s/2} \Gamma(s/2) \zeta(s)$ is the completed Riemann zeta function, invariant under $s\mapsto 1-s$.
The  Weyl reflection $w$ with respect to the root $\alpha_j$ acts by $\lambda\mapsto \lambda-(\lambda,\alpha_j)\alpha_j$. Using $\alpha_j=C_{ji}\lambda_i$ where  $\lambda_i$ are the fundamental weights and $C_{ij}$ is the Cartan matrix, the action  in weight basis and root basis is given by, respectively,
\be
w_j:\ \quad \sum_i x_i \lambda_i \to \sum (x_i - C_{ji} x_j )\, \lambda_i\ ,\quad
\sum_i y_i \alpha_i \to \sum (y_i - y_k C_{kj} \delta_{ij}) \alpha_i
\ee
The reflection coefficients satisfy the cocycle identity
\be
\label{eqcocycle}
M(w_1\cdot w_2,\lambda) = M(w_1, w_2\cdot \lambda)\, M(w_2,\lambda)\ .
\ee

An important characteristic of any automorphic form under $G$  is its constant term  with respect to the Borel subgroup $B$, i.e. its average under the action of the nilpotent subgroup $N$ of 
$B=AN$, generated by positive roots. For
$\cE^G(\lambda;g)$ it is given by Langlands' formula,
\be
\int_{N(\IZ)\backslash N(\IR)} \cE^G(\lambda;g\, n) \, \de n  
= \sum_{w\in W} M(w,\lambda) \, e^{\langle (w\cdot \lambda)+\rho, H(g)\rangle}\ .
\ee
This formula is consistent with the functional equation, thanks to \eqref{eqcocycle}.

The Langlands-Eisenstein series\eqref{defEisLang} satisfies the Laplace equation
\be
\Delta_G\, \cE^G(\lambda;g) = \frac12 \left( \langle\lambda,\lambda\rangle - \langle\rho,\rho\rangle\right)\, \cE^G(\lambda;g) 
\ee
Acting on monomials in the $t_i$'s, the Laplace-Beltrami operator reduces to 
\be
\Delta_G = \frac12 t^\rho\, C^{ij}\, t_i \pa_{t_i} t_j \pa_{t_j}  t^{-\rho} -  \frac12\langle\rho,\rho\rangle
\ee
where $t^\rho=\prod_{i=1}^r t_i$ and $C^{ij}$ is the inverse of the Cartan matrix.
Thus all constant terms are eigenmodes with the same eigenvalue. More generally,
Eisenstein series are eigenmodes of all invariant differential operators, with eigenvalue determined by the infinitesimal character $\lambda$.

For applications to BPS amplitudes in string theory, we are interested in the special case 
$\lambda=2s\lambda_{i_*}-\rho$, where $\lambda_{i_*}$ is a fundamental weight associated to a representation $\cR$. In that case, $\langle \lambda,\alpha_i\rangle=-1$ for all simple roots 
$\alpha_i$ with $i\neq i_*$, and the corresponding factor in $M(w,\lambda)$ vanishes. 
The sum over $B(\IZ)\backslash G(\IZ)$ then reduces to a sum
over $P_{i_*}(\IZ)\backslash G(\IZ)$, where $P_{i_*}$ is the maximal parabolic subgroup $G$ obtained by deleting the simple root $\alpha_{i_*}$ from the list of simple roots. 
We denote by $\cE^G_{\cR}(s;g)$
the resulting `maximal parabolic Eisenstein series', where $\cR$ is the finite-dimensional 
representation associated to the fundamental weight $\lambda_{i_*}$:
\be
\label{defEisLang}
\cE^G_{\cR}(s;g) 
= \sum_{\gamma\in P_\cR(\IZ) \backslash G(\IZ)} e^{\langle \lambda+\rho, H(\gamma g)\rangle}
\ee
The formula for the constant term with respect to the Borel subgroup reduces to
\be
\int_{N(\IZ)\backslash N(\IR)} \cE^G_\cR(s;g\, n) \, \de n  
= \sum_{w\in W/W_\cR} M(w,\lambda) \, e^{\langle (w\cdot \lambda)+\rho, H(g)\rangle}\ ,
\ee
where $W_\cR$ is the Weyl group of the Levi subgroup $L_{i_*}$ of $P_{i_*}$, or
equivalently the stabilizer of $\lambda_{i_*}$ in $W$. The sum therefore runs over
the Weyl orbit\footnote{The elements of $W/W_\cR$ are conveniently generated in LiE \cite{LiE} using the command 
{\tt for r row W$\_$orbit($\lambda_{i_*}$) do print(W$\_$word(r));print(",");   od}.}
 of $\lambda_{i_*}$. For $s=0$, all terms vanish except for $w\in W_\cR$, so that the constant term is equal to 1. In fact \cite[Thm 2.7]{Green:2010kv},
\be
\label{EG0}
\cE^{G}_{\cR}(0) = 1\ .
\ee

It is also of interest to extract the constant terms with respect to a maximal parabolic subgroup
$P_{i}$ (where $i$ could be the same as $i_*$). One way to do this is to scale  $t_j = u^{C^{ij}} t'_j$ and collect the constant terms with respect to the Borel subgroup
into powers of $u$ times constant terms of Langlands-Eisenstein series $\cE^{G'}_{\cR'}(s';g')$ associated to the Levi subgroup $G'$ of $P_i=G'\, N_i$. 
The result can be written as a sum over double cosets\footnote{The elements of $W_{i}\backslash W/W_\cR$ can be generated in LiE using {\tt double$\_$cosets$(L_i, L_{i_*})$}, where $L_i$
and $L_{i_*}$ are the  list of simple roots in the parabolic subgroups $P_i$ and $P_{i_*}$.}
\be
\int_{N_{i}(\IZ)\backslash N_{i}(\IR)} \cE^G_\cR(s;g\, n) \, \de n  
= \sum_{w\in W_{i}\backslash W/W_\cR} M(w,\lambda) \, e^{\langle [(w\cdot \lambda)+\rho]_{\parallel_i}, H(g)\rangle}\, \cE^{G_i}\left( (w\cdot\lambda)_{\perp i}, g' \right)
\ee
where $\lambda_{\parallel_i}, \lambda_{\perp i}$ denote the projection of $\lambda$ along and orthogonal to the fundamendal weight $\lambda_{i}$. Note that the Eisenstein series appearing on the r.h.s. are not necessarily associated to maximal parabolic subgroups.

In order to simplify the functional equations and analytic structure, it is convenient to 
consider the `completed' Langlands-Eisenstein series
\be
\cE^{G,\star} (\lambda;g)  =
L^G(\lambda)\, \cE^G(\lambda;g)
\ee
where
\be
L^G(\lambda)=
 \prod_{\alpha\in\Delta_+, w_L\cdot\alpha\in\Delta_-}   \zeta^\star(1+ \langle \lambda,\alpha \rangle)
\ee
where $w_L$ is the longest element in the Weyl group. $\cE^{G,\star} (\lambda;g)$ is then invariant under Weyl reflections,
\be
\cE^{\star,G}(\lambda;g)  =\cE^{\star,G}((w\cdot \lambda);g) \ .
\ee
Similarly, for maximal parabolic Eisenstein series, we denote 
\be
\cE^{\star,G}_{\cR}(s) = L^G_{\cR}(s)\, \cE^G_{\cR}(s)
\ee
where
\be
L^G_{\cR}(s) =  \prod_{\alpha\in\Delta_+, w_{\cR} \cdot \alpha\in\Delta_-}   \zeta^\star(1+ \langle \lambda,\alpha \rangle)
\ee
and $w_\cR$ is the longest element in the Weyl orbit of $\lambda_R$. 
 $\cE^{\star,G}_{\cR}(s)$ has a functional equation
\be
\cE^{\star,G}_{\cR}(s) = \cE^{\star,G}_{\cR'}(\kappa-s)\ ,\quad 
\kappa=\frac{ \langle\rho,\lambda_\cR\rangle}{ \langle\lambda_\cR,\lambda_\cR\rangle}
\ee
(where $\cR'$ is equal to $\cR$ or  to its image under an outer automorphism). 
Moreover, unlike $\cE^G_{\cR}(s)$,
its  meromorphic continuation in $s$ has only a finite number of poles.

In the physics literature, yet another normalization is commonly used\footnote{This is  denoted by ${\bf E}^{G}_{\cR,s}$ in \cite{Green:2010wi}, except for $G=SL(n), \cR=[010^{n-3}]$  where  an additional factor of $\zeta(2s-1)$ was inserted.}:
\be
E^{G}_{\cR,s} = 2 \zeta(2s) \cE^G_{\cR}(s)\ .
\ee
When $E^{G}_{\cR,s}$ has a pole at $s=s_0$, we denote by $\hat E^{G}_{\cR,s_0}$ the regularized
Eisenstein series, where the pole has been subtracted before taking the limit $s\to s_0$.
For $s=0$, one has, in view of \eqref{EG0},
\be
E^{G}_{\cR,0} = 2\zeta(0) = -1\ .
\ee

In the remainder of this section, we collect useful results about maximal parabolic Eisenstein series for $SL(d)$, $SO(d,d)$, $E_6$, $E_7$ and $E_8$. We label the simple roots using the same numbering as in LiE \cite{LiE}. Formulae for the constant terms can be found in the Mathematica
file {\tt EisensteinDefs.m} available from arXiv.

\subsection{$G=SL(d)$, $\cR=\Lambda^h[10^{d-2}]$}

For $h\leq d$, the representation of highest weight $[0^{h-1} 1 0^{d-h}]$ is the totally antisymmetric
tensor with $h$ indices. One has
\bea
\Delta_{SL(d)} \cE^{SL(d)}_{\Lambda^h[10^{d-2}]}(s) &=& \frac{hs(d-h)(2s-d)}{d} \cE^{SL(d)}_{\Lambda^h[10^{d-2}]}(s) \\
L^{SL(d)}_{\Lambda^h[10^{d-2}]}(s) &=&\prod_{k=1}^{\min(h,d-h)} \zetastar(2s+1-k)
\ ,\qquad \\ 
\cE^{\star,SL(d)}_{\Lambda^h[10^{d-2}]}(s)  &=& 
\cE^{\star,SL(d)}_{\Lambda^h[0^{d-2}1]}(\tfrac{d}{2}-s) 
\eea
For $h=1$ the only poles of $\cE^{\star SL(d)}_{[10^{d-2}]}(s)$ are at $s=0$ (originating from the normalizing factor) and $s=d/2$, with residue
\bea
{\rm Res}_{s=d/2} \cE^{\star SL(d)}_{[10^{d-2}]}(s) &=& \frac12
\eea
hence
\bea
{\rm Res}_{s=d/2} E^{SL(d)}_{[10^{d-2}],s} &=& \frac{\pi^{d/2}}{\Gamma(d/2)}\ ,\quad
\Delta_{SL(d)} \hat E^{SL(d)}_{[10^{d-2}],d/2} =  \frac{(d-1) \pi^{d/2}}{\Gamma(d/2)}
\eea
For $d=2$, the regularized Eisenstein series $\hat E^{SL(2)}_{[1],1}=\lim_{s\to 1}\left( 
E^{SL(2)}_{[1],s} - \frac{1}{s-1}\right)$ is given by the Kronecker limit formula,
\be
\hat E^{SL(2)}_{[1],1} = -\pi \log \tau_2 |\eta(\tau)|^4 + \mbox{cte}
\ee
where $\eta(\tau)$ is  the Dedekind eta function.

For $h=2$ the only poles of $\cE^{\star SL(d)}_{[010^{d-3}]}(s)$ are at $s=0$ (originating from the normalizing factor) and $s=d/2$, with residue
\bea
{\rm Res}_{s=d/2} \cE^{\star SL(d)}_{[010^{d-3}]}(s) &=& \frac12 \zeta^\star(2)
\eea
hence
\bea
{\rm Res}_{s=d/2} E^{SL(d)}_{[010^{d-3}],s} &=& \frac{(2\pi)^{d}}{24\Gamma(d-1)\zeta(d-1)}\ ,\\
\Delta_{SL(d)} \hat E^{SL(d)}_{[010^{d-3}],d/2} &=&  \frac{(2 \pi)^{d}}{12\Gamma(d-2)\zeta(d-1)} 
\eea
The decompactification limit $SL(d)\to SL(d-1)$ is obtained by setting 
\be
t_1=r^{d-1}\ ,\quad t_{2\leq i\leq d-1} = r^{d-i} t'_{i-1}
\ee
The Laplacian decomposes as
\be
\Delta_{SL(d)} = \Delta_{SL(d-1)} -\frac12 r\pa_r + \frac{1}{2d(d-1)} (r\pa_r)^2 
\ee
The constant term of the Eisenstein series with respect to the maximal parabolic
subgroup $P_1$ are given, for $h=1,2,d-2,d-1$, by  
\be
\begin{split}
\cE^{\star SL(d)}_{[10^{d-2}]}(s) \rightarrow& r^{d-2s}\, \cE^{\star SL(d-1)}_{[10^{d-3}0]}(s-\tfrac12) 
+ r^{2(d-1)s}\, \zeta^\star(2s)
\\
\cE^{\star SL(d)}_{[0^{d-2}1]}(s) \rightarrow& r^{2s}\, \cE^{\star SL(d-1)}_{[0^{d-3}1]}(s) 
+ r^{(d-1)(d-2s)}\, \zeta^\star(2s-d+1)
\\
\cE^{\star SL(d)}_{[010^{d-3}]}(s) \rightarrow& r^{2(d-2)s}\, \zeta^\star(2s-1)\, \cE^{\star SL(d-1)}_{[10^{d-3}]}(s) 
+r^{2(d-2s)} \cE^{\star SL(d-1)}_{[010^{d-4}]}(s-\tfrac12) \\
\cE^{\star SL(d)}_{[0^{d-3}10]}(s) \rightarrow& r^{(d-2)(d-2s)}\,\zeta^\star(2s-d+2)\, \cE^{\star SL(d-1)}_{[0^{d-3}1]}(s-\tfrac12) 
+r^{4s} \cE^{\star SL(d-1)}_{[0^{d-4}10]}(s) 
\end{split}
\ee

\subsection{$G=SO(d,d)$, $\cR=\Lambda^h[10^{d-1}]$}

For $h<d$,  the representation of highest weight $[0^{h-1} 1 0^{d-h+1}]$ is the totally antisymmetric
tensor with $h$ indices.  One has
\bea
\Delta_{SO(d,d)} \cE^{SO(d,d)}_{\Lambda^h[10^{d-1}]}(s) &=& h s (2s+ h+1-2d)\, \cE^{SO(d,d)}_{\Lambda^h[10^{d-1}]}(s) \nn\\
L^{SO(d,d)}_{\Lambda^h[10^{d-1}]}(s)&=&\zeta^\star(2s+h+1-d)
\prod_{k=0}^{h-1}\zeta^\star(2s-k)
\prod_{j=1}^{[h/2]} \zeta^\star(4s+2h+2-2d-2j)\nn\\
 \cE^{\star,SO(d,d)}_{\Lambda^h[10^{d-1}]}(s) &=& 
  \cE^{\star,SO(d,d)}_{\Lambda^h[10^{d-1}]}(d-\tfrac{h+1}{2} -s) 
\eea
In particular, for $1\leq h\leq 3$,
\be
\begin{split}
L^{SO(d,d)}_{[10^{d-1}]}(s)=&\zeta^\star(2s)\zeta^\star(2s+2-d)\ ,\\
L^{SO(d,d)}_{\Lambda^2[10^{d-1}]}(s)=&\zeta^\star(2s)\, \zeta^\star(2s-1)\, \zeta^\star(2s+3-d)\, \zeta^\star(4s+4-2d)\\
L^{SO(d,d)}_{\Lambda^3[10^{d-1}]}(s)=&\zeta^\star(2s)\, \zeta^\star(2s-1)\, \zeta^\star(2s-2)\, \zeta^\star(2s+4-d)\, \zeta^\star(4s+6-2d)
\end{split}
\ee
For $h=d$, we define
\be
\cE^{\star,SO(d,d)}_{\Lambda^d[10^{d-1}]}(s)=
 \prod_{k=0}^{d} \zeta^\star(2s+1-k)\, \left[ 
\cE^{\star,SO(d,d)}_{[0^{d-1}1]}(2s)+
\cE^{\star,SO(d,d)}_{[0^{d-2}10]}(2s) \right]
\ee
We do not attempt to define the series $\cE^{SO(d,d)}_{\Lambda^h[10^{d-1}]}$ for $h>d$.
The cases $d=1$ and $d=2$ are exceptional, 
\be
 \cE^{\star,SO(1,1)}_{[1]}(s) = \zeta^\star(2s)\, \zeta^\star(2s+1)\, (R^{2s}+R^{-2s}) \ .
\ee
\be
 \cE^{\star,SO(2,2)}_{[10]}(s) = \cE^\star(s;T)\, \cE^\star(s;U)\ ,\quad
 E^{SO(2,2)}_{V}(s) = \frac{1}{2\zeta(2s)} E(s;T)\, E(s;U)
\ee
For $d=3$, one has
\be
E^{SO(3,3)}_{[100],s}(t_1,t_2,t_3) = E^{SL(4)}_{[010],s}(t_2,t_1,t_3)
\ee

The only poles of $\cE^{\star SO(d,d)}_{[10^{d-1}]}(s)$ are at $s=0, \frac{d}{2}-1, \frac{d}{2}, d-1$, with the first two originating from the normalizing factor.
The residues at $s=\frac{d}{2},$ and $\frac{d}{2}-1$ are proportional to the minimal theta series.
The residues at $s=0$ and $s=d-1$ are constant,
\bea
{\rm Res}_{s=d-1} \cE^{\star SO(d,d)}_{[10^{d-1}]}(s) &=& \frac12 \zeta^\star(d-1)
\eea
hence
\bea
{\rm Res}_{s=d-1} E^{SO(d,d)}_{[10^{d-1}],s} &=& \frac{\pi^{d-1} \zeta^\star(d-1)}{\Gamma(d-1)\, \zeta^\star(d)}\ ,\quad\\ 
\Delta_{SO(d,d)} \hat E^{SO(d,d)}_{[10^{d-1}],d-1} &=&  
\frac{2\, \pi^{d-1} (d-1) \zeta^\star(d-1)}{\Gamma(d-1)\, \zeta^\star(d)}
\eea

Analysis of the constant terms shows that $\cE^{\star,SO(d,d)}_{\Lambda^2[10^{d-1}]}(s)$ has simple poles  at 
\be
\label{polesEisl2}
s=0,\tfrac12, \tfrac{d-3}{2}, \tfrac{d-2}{2},\tfrac{d-1}{2},\tfrac{d}{2}, d-2, d-\tfrac32
\ee
and double poles whenever these values coincide (except for $d=3$, $s=\tfrac12$ and $s=1$).
The first four values arise from poles of the normalizing factor. Similarly, $\cE^{\star,SO(d,d)}_{\Lambda^3[10^{d-1}]}(s)$ has simple poles at 
\be
\label{polesEisl3}
s=0,\tfrac12, 1, \tfrac{d-4}{2}, \tfrac{d-3}{2},\tfrac{d-1}{2},\tfrac{d}{2}, d-3, d-\tfrac52, d-2
\ee
and double poles whenever these values coincide. The first four values arise from poles of the normalizing factor.

The circle decompactification $SO(d,d)\to SO(d-1,d-1)$ is obtained by defining
\be
t_1 = R\ ,\quad t_{2\leq i\leq d-2} = R \,t'_{i-1}\ ,\quad t_{d-1}= R^{1/2} \, t'_{d-2}, \quad
t_{d-2} = R^{1/2} \, t'_{d-1}
\ee
The Laplacian decomposes as
\be
\Delta_{SO(d,d)}=\Delta_{SO(d-1,d-1)}+(1-d) R\pa_R +  \frac12 (R\pa_R)^2
\ee
The constant terms of the Eisenstein series with respect to the maximal parabolic
subgroup $P_1$ are given by 
\be
\begin{split}
 \cE^{\star,SO(d,d)}_{[10^{d-1}]}(s) \rightarrow &
  \zeta^\star(2s)\,  \zeta^\star(2s+2-d)\, R^{2s} 
 +  \zeta^\star(2s+1-d)\,  \zeta^\star(2s+3-2d)\, R^{2d-2-2s}
\\
 &+R\, \cE^{\star,SO(d-1,d-1)}_{[10^{d-2}]}(s-\tfrac12)\,
 \\
\cE^{\star,SO(d,d)}_{\Lambda^2[10^{d-1}]}(s) \to &  R^2\, \cE^{\star,SO(d-1,d-1)}_{\Lambda^2[10^{d-2}]}(s-\tfrac12) \\
&+ \zeta^\star(2s-1)\, \zeta^\star(4s+4-2d)\, R^{2s}\, \cE^{\star,SO(d-1,d-1)}_{[10^{d-2}]}(s) 
\\ 
& +  \zeta^\star(2s+5-2d)\, \zeta^\star(4s+3-2d)\, R^{2d-3-2s}\,  
\cE^{\star,SO(d-1,d-1)}_{[10^{d-2}]}(s-\tfrac12)
 \\
\cE^{\star,SO(d,d)}_{\Lambda^3[10^{d-1}]}(s) \to &  R^3\, \cE^{\star,SO(d-1,d-1)}_{\Lambda^3[10^{d-2}]}(s-\tfrac12) \\
&+ \zeta^\star(2s-2)\, R^{2s}\, \cE^{\star,SO(d-1,d-1)}_{\Lambda^2[10^{d-2}]}(s) 
\\ 
& +  \zeta^\star(2s+7-2d)\,  R^{2(d-2-s)}\,  
\cE^{\star,SO(d-1,d-1)}_{\Lambda^2[10^{d-2}]}(s-\tfrac12) 
\end{split}
\ee

The torus decompactification $SO(d,d)\to SL(d)$ is instead obtained by taking
\be
t_{1\leq i\leq d-2} = V^{i/d}\, t'_i, \quad t_{d-1}=V^{(d-2)/(2d)}\, t'_{d-1}\ ,\quad
t_d = V^{1/2}\ .
\ee
The Laplacian decomposes as
\be
\Delta_{SO(d,d)}=\Delta_{SL(d)}+\frac{d(1-d)}{2} V\pa_V +  \frac{d}{2} (V\pa_V)^2
\ee
The constant terms of the Eisenstein series with respect to the maximal parabolic
subgroup $P_{d}$ are given by 
\be
\begin{split}
  \cE^{\star,SO(d,d)}_{[10^{d-1}]}(s) \rightarrow & 
  V^{2s/d} \zeta^\star(2s+2-d)\, \cE^{\star,SL(d)}_{[10^{d-1}]}(s) \\
  &
 + V^{2-\frac{2s+2}{d}} \zeta^\star(2s+1-d)\, \cE^{\star,SL(d)}_{[0^{d-1}1]}(s+1-\tfrac{d}{2})
\\
\cE^{\star,SO(d,d)}_{\Lambda^2[10^{d-1}]}(s) \to &  
V^{4s/d}\, \zeta^\star(2s-d+3)\, \zeta^\star(4s-2d+4) 
\cE^{\star,SL(d)}_{\Lambda^2[10^{d-2}]}(s) \\
&+ V^{(4d-6-4s)/d} \zeta^\star(2s-d+1)\, \zeta^\star(4s-2d+3)\, 
\cE^{\star,SL(d)}_{\Lambda^2[0^{d-2}1]}(s+\tfrac{3-d}{2})
\\ 
& + V^{(2d-4)/d} \, \zeta^\star(2s-d+2)\, \cE^{\star,SL(d)}_{[s-\tfrac12,0^{d-3},s-\tfrac{d-3}{2}]}
\\
\cE^{\star,SO(d,d)}_{\Lambda^3[10^{d-1}]}(s) \to &  
V^{6s/d}\, \zeta^\star(2s-d+4)\, \zeta^\star(4s-2d+6) 
\cE^{\star,SL(d)}_{\Lambda^3[10^{d-2}]}(s) \\
&+ V^{6(d-2-s)/d} \zeta^\star(2s-d+1)\, \zeta^\star(4s-2d+3)\, 
\cE^{\star,SL(d)}_{\Lambda^3[0^{d-2}1]}(s+2-\tfrac{d}{2})
\\ 
& + V^{(2s+2d-6)/d}  \, \cE^{\star,SL(d)}_{[0,s-\tfrac12,0^{d-4},s-\tfrac{d}{2}+2]}
+ V^{(4d-10-2s)/d} \,  \cE^{\star,SL(d)}_{[s-1,0^{d-3},s-\tfrac{d}{2}+2]}
\end{split}
\ee
Notice that the terms on the last line in the equation for $h=2$ and $h=3$ are not maximal parabolic Eisenstein series.

\subsection{$G=SO(d,d)$, $\cR=[0^{d-1}1]$ and $\cR=[0^{d-2}10]$}
For the Eisenstein series attached to the spinor representations, one has 
\bea
\Delta_{SO(d,d)} \cE^{SO(d,d)}_{[0^{d-1}1]}(s) &=& \frac12 sd(s-d+1)\, \cE^{SO(d,d)}_{[0^{d-1}1]}(s) \\
L^{SO(d,d)}_{[0^{d-1}1]}(s)&=&\prod_{k=1}^{[d/2]}\zeta^\star(2s+2-2k)\ ,
\eea
The functional relation exchanges the two spinors when $d$ is odd,
\be
\begin{split}
\cE^{\star,SO(d,d)}_{[0^{d-1}1]}(s) = \cE^{\star,SO(d,d)}_{[0^{d-1}1]}(d-1-s) \qquad  d\, \mbox{even}\\
\cE^{\star,SO(d,d)}_{[0^{d-1}1]}(s) = \cE^{\star,SO(d,d)}_{[0^{d-2}10]}(d-1-s) \qquad  d\, \mbox{odd}
\end{split}
\ee
The series $\cE^{\star,SO(d,d)}_{[0^{d-1}1]}(s)$ has first order poles at $s=0,1,2,\dots, d-1$ (except at $\frac{d-1}{2}$, if this happens to be integer). The poles at $s=0,1,\dots, [d/2]-1$ originate from the normalizing factor. It is important to note that $\cE^{\star,SO(d,d)}_{[0^{d-1}1]}(s)-\cE^{\star,SO(d,d)}_{[0^{d-2}10]}(s)$ is an entire function of $s$.
$E^{SO(d,d)}_{[0^{d-1}1],s}(t_i)$ and $E^{SO(d,d)}_{[0^{d-2}10],s}(t_i)$  are finite and coincide
when $s$ takes any integer or half-integer value from $s=1$ to $s=\tfrac{d-1}{2}$.
They have first order poles at integer values in the interval $\tfrac{d-1}{2}<s\leq d-1$.

For $d=2$, the Grassmannian $SO(2,2)/SO(2)\times SO(2)$ decomposes into the product
of two Poincar\'e upper half planes, parametrized by complex moduli $T$ and $U$. We have
\be
\cE^{\star,SO(2,2)}_{[10]}(s) = \cE^{\star,SL(2)}_{[1]}(s;T)\ ,\quad 
\cE^{\star,SO(2,2)}_{[01]}(s) =\cE^{\star,SL(2)}_{[1]}(s;U)\ .
\ee
For $d=4$, triality relates the vector and spinor Eisenstein series at different points,
\be
E^{SO(4,4)}_{[0001],s}(t_1,t_2,t_3,t_4)= E^{SO(4,4)}_{[0010],s}(t_1,t_2,t_4,t_3)=
E^{SO(4,4)}_{[1000],s}(t_4,t_2,t_3,t_1)
\ee
For $s=1$, triality further equates the vector and spinor Eisenstein series at the same point,
\be
E^{SO(4,4)}_{[1000],1}(t_1,t_2,t_3,t_4)=E^{SO(4,4)}_{[0010],1}(t_1,t_2,t_3,t_4)=E^{SO(4,4)}_{[0001],1}(t_1,t_2,t_3,t_4)
\ee
The constant terms with respect to maximal parabolic subgroups $P_1$ and $P_d$
are 
\be
\begin{split}
 \cE^{\star,SO(d,d)}_{[0^{d-1}1]}(s) \rightarrow &
 R^{s} \ \cE^{\star,SO(d-1,d-1)}_{[0^{d-2}1]}(s)\\
 & + R^{d-1-s}  \cE^{\star,SO(d-1,d-1)}_{[0^{d-3}10]}(s-1)\quad (d \ {\rm odd})
 \\
  \cE^{\star,SO(d,d)}_{[0^{d-1}1]}(s) \rightarrow &
 R^{s} \zeta^\star(2s+2-d)\, \cE^{\star,SO(d-1,d-1)}_{[0^{d-2}1]}(s)\\
 &+ R^{d-1-s} \zeta^\star(2s+1-d)\,  \cE^{\star,SO(d-1,d-1)}_{[0^{d-2}1]}(s-1)\quad (d \ {\rm even})
\\
 \cE^{\star,SO(d,d)}_{[0^{d-1}1]}(s) 
\to & \sum_{k=0\dots d \atop k\, {\rm even}} 
V^{\frac{k(k-1)+s(d-2k)}{d}}\, L_k(s)\, \cE^{\star,SL(d)}_{\Lambda^{d-k}[10^{d-2}]}(s-\tfrac{k-1}{2})
\\
 \cE^{\star,SO(d,d)}_{[0^{d-2}10]}(s) 
\to & \sum_{k=0\dots d \atop k\, {\rm odd}} 
V^{\frac{k(k-1)+s(d-2k)}{d}}\, L_k(s)\, \cE^{\star,SL(d)}_{\Lambda^{d-k}[10^{d-2}]}(s-\tfrac{k-1}{2})
\end{split}
\ee
where
\be
L_k(s)=\prod_{\ell=k+1}^{[d/2]} \zeta^\star(2s+2-2\ell)\,
\prod_{\ell=d-k+1}^{[d/2]} \zeta^\star(2s+2\ell+1-2d)
\ee

\subsection{$E_6$}

 $\cE^{\star,E_6}_{[100000]}(s)$ (corresponding to one of the two irreducible representations of dimension 27) has normalizing factor
\be
L^{E_6}_{[100000]}(s) = \zeta^\star(2s)\, \zeta^\star(2s-3)\ ,
\ee
and simple poles at $s=0,3/2,9/2,6$. The constant term with respect to $P_6$ yields the decompactification limit
\be
\begin{split}
\cE^{\star,E_6}_{[100000]}(s)\to & 
R^{\tfrac43 s}\, \cE^{SO(5,5),\star}_{[10000]}(s)+  
R^{5-\tfrac23 s}\, \cE^{SO(5,5),\star}_{[00010]}(s-\tfrac32)\\
&+ R^{-\tfrac83(s-6)}\, \zetastar(2s-8)\, \zetastar(2s-11)\ ,
\end{split}
\ee
while the constant term with respect to $P_1$ yields the weak coupling limit,
\be
\begin{split}
\cE^{\star,E_6}_{[100000]}(s)\to & g_5^{-8s/3}\, \zetastar(2s)\, \zetastar(2s-3)
+g_5^{-3-\frac23 s} \cE^{SO(5,5),\star}_{[00001]}(s-\tfrac12)
+g_5^{\frac43(s-6)}\, \cE^{SO(5,5),\star}_{[10000]}(s-2)\
\end{split}
\ee
The functional equation is 
\be
\cE^{\star,E_6}_{[100000]}(s)= \cE^{\star,E_6}_{[000001]}(6-s)\ .
\ee

\subsection{$E_7$}

$\cE^{\star,E_7}_{[1000000]}(s)$ (corresponding to the irreducible representation of dimension 133) has normalizing factor
\be
L^{E_7}_{[1000000]}(s) = \zeta^\star(2s)\, \zeta^\star(2s-3) \, \zeta^\star(2s-5)  \, \zeta^\star(4s-16)\ ,
\ee
is invariant under $s\mapsto \tfrac{17}{2}-s$, and has 
simple poles at $s=0,3/2,5/2,4,9/2,6,7,17/2$. The constant term with respect to $P_7$ yields the  decompactification limit
\be
\begin{split}
\cE^{\star,E_7}_{[1000000]}(s) \to &
R^{2s}\,  \zeta^\star(2s-5)\,  \zeta^\star(4s-16)\, \cE^{\star,E_6}_{[100000]}(s) \\
&+ R^{17-2s}   \zeta^\star(2s-11)\, \zeta^\star(4s-17)\, \cE^{\star,E_6}_{[000001]}(s-\tfrac{5}{2}) \\ 
&+R^{6} \, \cE^{\star,E_6}_{[010000]}(s-\tfrac{3}{2})
\end{split}
\ee
while the constant term with respect to $P_1$ yields the  weak coupling limit
\be
\begin{split}
\cE^{\star,E_7}_{[1000000]}(s) \to &
g_4^{-8}\, \, \cE^{\star,D_6}_{[010000]}(s-2) \\
&+ g_4^{-4s}     \zeta^\star(2s)\,   \zeta^\star(2s-3)\, \zeta^\star(2s-5)\, \zeta^\star(4s-16)\\
&+ g_4^{4s-34}     \zeta^\star(2s-6)\,   \zeta^\star(2s-8)\, \zeta^\star(2s-10)\, \zeta^\star(4s-17)\\
&+g_4^{-2s-1} \, \zeta^\star(4s-16)\, \cE^{\star,D_6}_{[000001]}(s-\tfrac12) \\
&+g_4^{2s-18} \, \zeta^\star(4s-17)\, \cE^{\star,D_6}_{[000001]}(s-3) \
\end{split}
\ee

 $\cE^{\star,E_7}_{[0000001]}(s)$ (corresponding to the irreducible representation of 
dimension 56) has normalizing factor
\be
L^{E_7}_{[0000001]}(s) = \zeta^\star(2s)\, \zeta^\star(2s-4) \, \zeta^\star(2s-8) \ ,
\ee
is invariant under $s\mapsto 9-s$, has 
simple poles at $s=0,2,4,5,7,9$. The constant term with respect to $P_7$ is
\be
\begin{split}
\cE^{\star,E_7}_{[0000001]}(s)\to& 
 R^{3s}\, \zetastar(2s)\, \zetastar(2s-4)\, \zetastar(2s-8)\\
 &+R^{s+1}\, \zetastar(2s-8)\,  \cE^{E_{6},\star}_{[000001]}(s-\tfrac12)\\
&+R^{10-s} \, \zetastar(2s-9)\, \cE^{E_{6},\star}_{[100000]}(s-\tfrac52)\\
&+R^{3(9-s)} \zetastar(2s-9)\, \zetastar(2s-13) \, \zetastar(2s-17)   
\end{split}
\ee
while the constant term with respect to $P_1$ is
\be
\begin{split}
\cE^{\star,E_7}_{[0000001]}(s)\to& g_4^{-2s}\, \zetastar(2s-8)\,  \cE^{\star,E_6}_{[100000]}(s) 
+ g_4^{2s-18}\, \zetastar(2s-9)\,  \cE^{\star,E_6}_{[100000]}(s-4)  \\
&+ g_4^{-6}\,  \cE^{\star,E_6}_{[000010]}(s-2) 
\end{split}
\ee

\subsection{$E_8$}

$\cE^{\star,E_8}_{[00000001]}(s)$ (corresponding to the irreducible representation of dimension 248)  has normalizing factor
\be
L^{E_8}_{[00000001]}(s) = \zeta^\star(2s)\, \zeta^\star(2s-5) \, \zeta^\star(2s-9) \,  \zeta^\star(4s-28) \ ,
\ee
is invariant under $s\mapsto \tfrac{29}{2}-s$, and has simple 
poles at $0,\tfrac52,\tfrac92, 7,\tfrac{15}{2},10,12,\tfrac{29}{2}$. The constant term with respect to
$P_8$ yields the decompactification limit
\be
\begin{split}
\cE^{\star,E_8}_{[00000001]}(s) \to&  
R^{4s}\, \zetastar(2s)\, \zetastar(2s-5)\, \zetastar(2s-9)\, \zetastar(4s-28)
\\
&+ R^{2s+1} \, \zetastar(4s-28)\,  \cE^{\star,E_7}_{[0000001]}(s-\tfrac12)\\
&+  R^{12}\, \cE^{\star,E_7}_{[1000000]}(s-3)\\
&+ R^{2(15-s)} \, \zetastar(4s-29)\,  \cE^{\star,E_7}_{[0000001]}(s-5)
\\
&+ R^{2(29-2s)} \zetastar(2s-19)\, \zetastar(2s-23) \, \zetastar(2s-28) \, \zetastar(4s-29) 
\end{split}
\ee
while the constant term with respect to $P_1$ gives the weak coupling limit
\be
\begin{split}
\cE^{\star,E_8}_{[00000001]}(s) \to&  
g_3^{-20} \,  \cE^{\star,D_7}_{[0100000]}(s-\tfrac92) \\
& + g_3^{-4s}\, \zetastar(2s-9)\, \zetastar(4s-28)\, \cE^{\star,D_7}_{[1000000]}(s)\\
& + g_3^{4s-58}\, \zetastar(2s-9)\, \zetastar(4s-29)\, \cE^{\star,D_7}_{[1000000]}(s-\tfrac{17}{2})\\
& + g_3^{-7-2s}\, \zetastar(4s-28)\, \cE^{\star,D_7}_{[0000001]}(s-\tfrac52)\\
& + g_3^{2s-36}\, \zetastar(4s-29)\, \cE^{\star,D_7}_{[0000010]}(s-6)
\end{split}
\ee

 $\cE^{\star,E_8}_{[10000000]}(s)$ (corresponding to the irreducible representation of dimension 3875)  has normalizing factor 
\be
L^{E_8}_{[10000000]}(s) = \zeta^\star(2s)\, \zeta^\star(2s-3) \, \zeta^\star(2s-5) \, \zeta^\star(2s-6) \, \zeta^\star(2s-9)  \, \zeta^\star(4s-16)\, \zeta^\star(4s-22) \ ,
\ee
is invariant under $s\mapsto \tfrac{23}{2}-s$, and has simple 
poles at $0,\tfrac32, \tfrac52, 3,4,\tfrac92,5,\tfrac{11}{2},6,\tfrac{13}{2},7,\tfrac{15}{2},\tfrac{17}{2},9,10,\tfrac{23}{2}$.   The constant term with respect to $P_8$ is
\be
\begin{split}
\cE^{\star,E_8}_{[10000000]}(s) \to &
R^{4s}\,  \zeta^\star(2s-6)\,  \zeta^\star(2s-9)\,  \zeta^\star(4s-22)\, \cE^{\star,E_7}_{[1000000]}(s) \\
&+ R^{46-s}  \zeta^\star(2s-13)\,  \zeta^\star(2s-16)\,  \zeta^\star(4s-23)\, 
\cE^{\star,E_7}_{[1000000]}(s-3)\\
&+R^{2s+7}  \zeta^\star( 4s-22)\, \cE^{\star,E_7}_{[0100000]}(s-\tfrac{3}{2}) \\
&+R^{30-2s}   \zeta^\star(4s-23)\, \cE^{\star,E_7}_{[0100000]}(s-3) \\ 
&+R^{18} \,  \zeta^\star(2s-11)\, \cE^{\star,E_7}_{[0000010]}(s-\tfrac{5}{2})
\end{split}
\ee
while the constant term with respect to $P_1$ is
\be
\begin{split}
\cE^{\star,E_8}_{[10000000]}& (s ) \to g_3^{-8s} \, \zetastar(2s)\,  \zetastar(2s-3)\,   \zetastar(2s-5)\,  \zetastar(2s-6)\,  \zetastar(2s-9)\,  \\
& \times \zetastar(4s-16)\,  \zetastar(4s-22)
+  g_3^{8s-92} \, \zetastar(2s-13)\,  \zetastar(2s-16)\,  \\
&\times \zetastar(2s-17)\,  \zetastar(2s-19)\,  \zetastar(2s-22)\,  \zetastar(4s-23)\,  \zetastar(4s-29)
 \\
& +g_3^{-4(s+2)}\,  \zetastar(2s-9)\,  \zetastar(4s-22)\, \cE^{\star,D_7}_{[0010000]}(s-2)
\\
&+g_3^{-1-6s}\, \zetastar(2s-6)\,  \zetastar(2s-9)\,  \zetastar(4s-16)\,  \zetastar(4s-22)\,
\cE^{\star,D_7}_{[0000010]}(s-\tfrac12)
\\
& + g_3^{-70+6s}\, \zetastar(2s-13)\,  \zetastar(2s-16)\,  \zetastar(4s-23)\,  \zetastar(4s-29)\,
\cE^{\star,D_7}_{[0000001]}(s-5)\\
&+g_3^{4s-54}\,  \zetastar(2s-13)\,  \zetastar(4s-23)\, \cE^{\star,D_7}_{[0010000]}(s-\tfrac92)
\\
&+g_3^{-28}\,  \zetastar(2s-9)\,  \zetastar(4s-20)\, \cE^{\star,D_7}_{[0001000]}(s-\tfrac72)
\\
&+g_3^{-34}\,  \zetastar(2s-6)\,  \zetastar(2s-9)\, \zetastar(2s-11)\,   \zetastar(2s-13)\,  \zetastar(2s-16)\,
 \cE^{\star,D_7}_{[1000000]}(2s-\tfrac{17}{2})
 \\
& +g_3^{-18-2s}\,  \cE^{\star,D_7}_{[1000010]}(s-3)
+g_3^{2s-41}\,  \cE^{\star,D_7}_{[s-5,00000,s-\tfrac72]} 
\end{split}
\ee
Note that the last term is not a maximal parabolic Eisenstein series.

\section{Weak coupling and large radius expansions  for $D\geq 3$ \label{sec_alldim}}
 
 In this appendix we provide the weak coupling and large radius expansions of $\cR^4, D^4\cR^4$ and $D^6 \cR^4$ couplings in all dimensions $D\geq 3$. These expansions agree by and large with the results in \cite{Green:2010wi,Pioline:2010kb,Green:2010kv,Green:2011vz} when available, except for certain important numerical coefficients. For $\cR^4$ and $D^4\cR^4$ couplings, we also recall the known expressions in terms of Langlands-Eisenstein series of the U-duality group. 
Detailed computations can be found in the Mathematica worksheet {\tt d6r4eisenstein.nb} 
available on arXiv.

\subsection{$D=10, d=0$}

In ten-dimensional type IIB string theory, the moduli space $SL(2,\IR)/U(1)$ is parame\-trized
by the axiodilaton $\tau= C_0+\frac{\I}{g}$, identified under $SL(2,\IZ)$ S-duality. In contrast,
 ten-dimensional type IIA string theory has only one real modulus, the string coupling, and trivial
 S-duality group. The perturbative contributions to $\cR^4$, $D^4 \cR^4$ and $D^6 \cR^4$
 are identical in type IIA and type IIB, as they do not receive contributions from odd-odd spin structures. Type IIA has no D-instantons, accordingly these couplings do not receive any non-perturbative corrections. The decompactification to $D=11$ can be analyzed using 
 $g l_s =R_{11}$, $g_s l_s^3= l_M^3$.

\subsubsection{$\cR^4$}

The $\cR^4$ coupling in type IIB theory in 10 dimensions is given by \cite{Green:1997tv}
\be
 \cE^{(0)}_{(0,0)} = E^{SL(2)}_{[1],3/2} \ .
\ee
At weak coupling, this produces the expected tree-level and one-loop corrections,
along with an infinite series of D-instanton corrections,
\be
g^{-1/2}\, \cE^{(0)}_{(0,0)} = \frac{2\zeta(3)}{g^2} + 4 \zeta(2)  + {\rm n.p.}
\ee
In type IIA string theory, only the two perturbative terms are present.  
Under decompactification to 11 dimensions, the tree-level $\cR^4$ term is 
suppressed as $1/R_{11}^2$, while the one-loop $\cR^4$ term scales as $R_{11}/l_M^3$, so corresponds to a one-loop correction in the 11-dimensional local effective action.

\subsubsection{$D^4 \cR^4$}

The $D^4 \cR^4$ coupling in type IIB theory in 10 dimensions is given by \cite{Green:1999pu}
\be
 \cE^{(0)}_{(1,0)} = E^{SL(2)}_{[1],5/2} 
\ee
At weak coupling, this produces the expected tree-level, one-loop (vanishing) and two-loop corrections \cite{Green:1999pv,D'Hoker:2005ht},
along with an infinite series of D-instanton corrections,
\be
g^{1/2}\, \cE^{(0)}_{(1,0)} = \frac{\zeta(5)}{g^2} + 0+ \frac43 \zeta(4) g^2  + {\rm n.p.}
\ee
In type IIA string theory, under decompactification to 11 dimensions, the tree-level $D^4 \cR^4$ term is suppressed as $l_M^6/R_{11}^4$, while the two-loop $D^4\cR^4$ term scales as $R_{11}^2$, corresponding to the first term in an infinite series of terms which sum up to a non-local term in $D=11$.  Thus, there is no $D^4\cR^4$ term in the local action in $D=11$, for the same reason
that a term $\cR^6$ cannot appear \cite{Russo:1997mk}.

\subsubsection{$D^6 \cR^4$}

The perturbative corrections to the $D^6 \cR^4$ coupling in type IIB theory in 10 dimensions have been computed in \cite{Green:1999pv,Green:2005ba,Green:2008uj,D'Hoker:2013eea}. Based on an extensive analysis of loop amplitudes in 11D supergravity \cite{Green:1999pu,Green:2008bf,D'Hoker:2014gfa}, it is believed that there are no perturbative corrections beyond 3-loop: 
\be
g\, \cE^{(0)}_{(0,1)} = \frac{2\zeta^2(3)}{3g^2} + \tfrac43 \zeta(2)\zeta(3)+
\tfrac85 \zeta^2(2) g^2 + \tfrac{4}{27} \zeta(6) g^4  + {\rm n.p.}
\ee
A non-perturbative completion satisfying the Poisson equation \eqref{eisenthree} was proposed in \cite{Green:2005ba}, and the resulting non-perturbative effects were  analyzed in 
\cite{Green:2014yxa}. 

In type IIA string theory, only the four perturbative terms are present.
Under decompactification to 11 dimensions, the tree-level and one-loop $D^6 \cR^4$ terms are suppressed as $l_M^9/R_{11}^5$ and $l_M^6/R_{11}^2$, respectively, while the two-loop term scales as $l_M^3 R_{11}$, corresponding to a two-loop correction in 
the 11-dimensional local effective action \cite{Green:2005ba}. The three loop term 
scales as $R_{11}^4$, corresponding again to the first term in an infinite series of terms which sum up to a non-local term in $D=11$.

\subsection{$D=9, d=1$}
In type II string theory compactified on $S^1$,
the moduli space is $\IR^+\times SL(2,\IR)/U(1)$, identified under $SL(2,\IZ)$ leaving the first factor inert. The first factor is parametrized by 
\be
\nu=\left(\frac{r}{l_s}\right)^{7/4}\sqrt{g_9} = \left(\frac{l_{10B}}{\tilde r}\right)^2\ ,
\ee
where $r$ is the radius of the 
type IIA circle, $g_9$ is the string coupling in $D=9$, while $\tilde r=l_s^2/r$ is the radius 
of the type IIB circle and  $l_{10B}=l_s g^{1/4}$ is the Planck length in ten-dimensional 
type IIB. The second factor is parametrized by the type IIB axiodilaton 
\be
\tau= C_0+\frac{\I}{g} = C_1+\I \frac{\sqrt{r/l_s}}{g_9} \ .
\ee
The decompactification to ten-dimensional 
type IIB corresponds to $\cV\to 0$ keeping $\tau$ finite. The decompactification to 
ten-dimensional type IIA theory instead corresponds to $\cV,\tau_2\to\infty$ keeping
$\tau_2^3/\cV=g_{10A}^{-7/2}$ fixed.

\subsubsection{$\cR^4$}
The exact $\cR^4$ coupling is given by \cite{Green:1997di,Green:1997as}
\be
\cE^{(1)}_{(0,0)} = \nu^{-3/7}\, E^{SL(2)}_{[1],3/2}(\tau) + 4 \zeta(2)\, \nu^{4/7}
\ee
The two contributions can be separated by considering different kinematics \cite{Kiritsis:1997em}. 
At weak coupling, this Ansatz produces the expected tree-level and one-loop contributions,
along with an infinite series of D-instanton corrections,
\be
g_9^{-2/7}\, \cE^{(1)}_{(0,0)} = \frac{2\zeta(3)}{g_9^2} + 4 \zeta(2) 
\left(\frac{r}{l_s}+ \frac{l_s}{r} \right)
 + {\rm n.p.}
\ee
Decompactifying from $D=9$ to $D=10B$, with $r_1=\tilde r$, one has, in agreement with
\eqref{decompR4}, 
\be
\cE^{(1)}_{(0,0)} \rightarrow \left(\frac{r_1}{l_{10B}}\right)^{6/7} \cE^{(0)}_{(0,0)} 
+ 4\zeta(2)\,  \left(\frac{r_1}{l_{10B}}\right)^{-8/7} 
\ee

\subsubsection{$D^4 \cR^4$}
The exact $D^4\cR^4$ coupling is given by \cite{Green:1999pu}
\be
\cE^{(1)}_{(1,0)} = \frac12 \nu^{-5/7}\, E^{SL(2)}_{[1],5/2} 
+ \frac{2}{15} \zeta(2)\, \nu^{9/7}\, E^{SL(2)}_{[1],3/2} 
+ \frac{4\zeta(2)\zeta(3)}{15} \nu^{-12/7}
\ee
This produces the expected tree-level, one-loop and two-loop contributions
\be
g_9^{6/7}\, \cE^{(1)}_{(1,0)} = \frac{\zeta(5)}{g_9^2} 
+ \frac{4}{15} \zeta(2) \zeta(3) \left( \frac{r^3}{l_s^3}+ \frac{l_s^3}{r^3}\right)
+ \frac43 \zeta(4)\, g_9^2 \left(\frac{r^2}{l_s^2}+\frac{l_s^2}{r^2}\right) + {\rm n.p.}
\ee
Decompactifying from $D=9$ to $D=10$ \cite[4.8]{Green:2010wi}, one has, in agreement with
\eqref{decompD4R4},
\be
\cE^{(1)}_{(1,0)} \rightarrow \left(\frac{r_1}{l_{10B}}\right)^{10/7}
 \left( \cE^{(0)}_{(1,0)} 
+ \tfrac{2}{15} \zeta(2) \,  \left(\frac{r_1}{l_{10B}}\right)^{-4}\, 
  \cE^{(0)}_{(0,0)} +\tfrac{4}{15}\zeta(2)\zeta(3) \left(\frac{r_1}{l_{10B}}\right)^2  \right)
\ee

\subsubsection{$D^6 \cR^4$}

The exact $D^6 \cR^4$ coupling is believed to be given by \cite{Green:2010wi}
\be
\begin{split}
 \cE^{(1)}_{(0,1)} = \nu^{-\frac67}   \cE_{(0,1)}^{(0)} 
 + \frac23\zeta(2) \nu^{\frac17} E^{SL(2)}_{[1]3/2}
 + \frac2{63}\zeta(2) \nu^{\frac{15}7} E^{SL(2)}_{[1]5/2}
 +\frac4{63}\zeta(2)\zeta(5) \nu^{-\frac{20}7}
 +\frac85\zeta(2)^2 \nu^{\frac{8}7}
 \end{split}
\ee
where the first term is proportional to the $D^6 \cR^4$ coupling in 10 dimensions.
At weak coupling, this produces the expected perturbative terms up to 
three-loops,
\be
\begin{split}
g_9^{10/7}\, \cE^{(1)}_{(0,1)} = &  \frac{2\zeta(3)^2}{3 g_9^2} +
\left[ \frac43 \zeta(2)\zeta(3) \, (r + \frac{1}{r}) + \frac{4}{63} \zeta(2) \zeta(5) \, (r^5 +\frac{1}{r^5} )
\right]\\
& + 8 \zeta(2)^2   \left[ \frac13 + \frac15 
 \left(r^2 + \frac{1}{r^2} \right) \right]\, g_9^2 + \frac{4}{27} \zeta(6) \, (r^3 + \frac{1}{r^3}) \, g_9^4 + {\rm n.p.}
 \end{split}
\ee
Decompactifying from $D=9$ to $D=10$, one has, in agreement with \eqref{decompD6R4},
\be
\begin{split}
\cE^{(1)}_{(0,1)} \rightarrow  & \left(\frac{r_1}{l_{10B}}\right)^{12/7} \left( \cE^{(0)}_{(0,1)} 
+ \frac{4}{63}\zeta(2)\, \left(\frac{r_1}{l_{10B}}\right)^{-6} \, \cE^{(0)}_{(1,0)}
+ \frac23 \zeta(2)  \left(\frac{r_1}{l_{10B}}\right)^{-2}\, \cE^{(0)}_{(0,0)} \right.\\
&\left.
+\frac{8}{5}\zeta^2(2) \left(\frac{r_1}{l_{10B}}\right)^{-4} + \frac{2\pi^2}{189}\zeta(5) \left(\frac{r_1}{l_{10B}}\right)^4 \right)
\end{split}
\ee

\subsection{$D=8, d=2$}

In type II string theory compactified on $T^2$, the moduli space is a product $SL(3)/SO(3)\times SL(2)/U(1)$, identified under $E_3=SL(3,\IZ)\times SL(2,\IZ)$. In type IIA, the first factor parametrizes the dilation, K\"ahler modulus $T$ and RR axions, while the second corresponds to the complex modulus $U$ of the two-torus. In type IIB, the role of $T$ and $U$ is exchanged. 

\subsubsection{$\cR^4$}
The exact $\cR^4$ coupling is given by a linear combination of two Eisenstein series \cite{Kiritsis:1997em}
\be
\cE^{(2)}_{(0,0)} =\hat E^{SL(3)}_{[10],3/2} + 2 \hat E^{SL(2)}_{[1],1}(U)
\ee
The hat indicates that the simple poles at $s=3/2$ and $s=1$, respectively, have been subtracted.
The two Eisenstein series can be disentangled by considering different kinematics in the four-graviton scattering. They produce the expected tree-level and one-loop contributions,
\be
\cE^{(2)}_{(0,0)} = \frac{2\zeta(3)}{g_8^2} +  2(  \hat E^{SL(2)}_{[1],1}(T)  + \hat E^{SL(2)}_{[1],1}(U) )
+\frac{4\pi}{3} \log g_8
 + {\rm n.p.}
\ee
Decompactifying from $D=8$ to $D=9$, one obtains, in agreement with \eqref{decompR4}
\be
\label{ER48R}
\cE^{(2)}_{(0,0)} \rightarrow \frac{r_2}{l_9} \cE^{(1)}_{(0,0)} - \frac{14\pi}{3} \log \frac{r_2}{l_9}\ ,\quad
\ee
This is consistent with the non-analytic term $\frac{4\pi}{3}\, \log g_8$ in \eqref{Enonan},
using $l_9=l_s g_9^{2/7}$.

\subsubsection{$D^4 \cR^4$}
The exact $D^4\cR^4$ coupling is given by \cite{Basu:2007ru,Green:2010wi} 
\be
\cE^{(2)}_{(1,0)} =\frac12 E^{SL(3)}_{[10],5/2} - 4  E^{SL(2)}_{[1],2}(U) \, E^{SL(3)}_{[10],-1/2}
\ee
At weak coupling, this produces the expected tree-level, one-loop and two-loop contributions,
\be
g_8^{4/3}\, \cE^{(2)}_{(1,0)} = \frac{\zeta(5)}{g_8^2} + 
\frac{2}{\pi^3} E^{SL(2)}_{[1],2}(T)  E^{SL(2)}_{[1],2}(U) 
+ \frac23 g_8^2 (  E^{SL(2)}_{[1],2}(T)  + E^{SL(2)}_{[1],2}(U) )
 + {\rm n.p.}
\ee
Decompactifying from $D=8$ to $D=9$, one has, in agreement with \eqref{decompD4R4},
\be
\cE^{(2)}_{(1,0)} \rightarrow \left(\frac{r_2}{l_{9}}\right)^{5/3} \left( \cE^{(1)}_{(1,0)} 
+ \tfrac{1}{\pi} \zeta(3)\,  \left(\frac{r_2}{l_{9}}\right)^{-3} 
  \cE^{(1)}_{(0,0)} +\tfrac{4\pi}{45}\zeta(4) \left(\frac{r_2}{l_{9}}\right)^3  \right)
\ee

\subsubsection{$D^6 \cR^4$}

The exact $D^6\cR^4$ coupling was proposed in \cite{Green:2010wi}, building on \cite{Basu:2007ck}: 
\be
\label{D6R48}
\begin{split}
\cE^{(2)}_{(0,1)} = &  \cE_{(0,1)}^{SL(3)} + \cE_{(0,1)}^{SL(2)}+ \frac13 \hat E^{SL(3)}_{[10],3/2}\,\hat E^{SL(2)}_{[1],1} 
+ \frac{\pi}{36}  \hat E^{SL(3)}_{[10],3/2} + \frac{\pi}{9} \hat E^{SL(2)}_{[1],1} 
 + \frac{\zeta(2)}{9} \\
 &+ \frac{40}{9} E^{SL(3)}_{[10],-3/2}\, E^{SL(2)}_{[1],3}
\end{split}
\ee
Here, $\cE_{(0,1)}^{SL(2)} $ is the solution to
\be
(\Delta_U-12) \, \cE_{(0,1)}^{SL(2)}  = -4 \left[ \hat E^{SL(2)}_{[1],1}(U) \right]^2
\ee
which behaves in the limit  $U_2\to\infty$ as 
\be
\begin{split}
6\, \cE_{(0,1)}^{SL(2)} =&\frac{\pi^2}{180}(65-20\pi U_2+48\pi^2 U_2^2)+\frac{\zeta(3)\zeta(5)}{\pi U_2^3} \\
&
-2\zeta(2) \, (4\pi U_2-6 \log U_2+1)\, \log U_2  + \cO(e^{-U_2})
\end{split}
\ee
while $\cE_{(0,1)}^{SL(3)}$ is a solution to 
\be
(\Delta_{SL(3)} - 12) \cE_{(0,1)}^{SL(3)} = - \left( \hat E^{SL(3)}_{[10],3/2}  \right)^2 
\ee
behaving in the limit $g_8\to 0$ as 
\be
\begin{split}
g_8^{2}\, \cE_{(0,1)}^{SL(3)} = & \frac{2\zeta(3)^2}{3g_8^2} +  \frac23\zeta(3)   \hat E^{SL(2)}_{[1],1}(T) + \left( f(T) + \frac{\pi}{18} \hat E^{SL(2)}_{[1],1}(T) \right) g_8^2 + \frac{2}{27} 
E^{SL(2)}_{[1],3}(T)\\ 
+ & \frac{4\pi^2}{27} g_8^2 \log^2 g_8
+ 2 \log g_8 \left[ \frac{2\pi \zeta(3)}{9} + g_8^2 \left( \frac{6\pi}{27}  \hat E^{SL(2)}_{[1],1}(T) \ + 
\frac{\pi^2}{27} \right) \right] + \frac{7\pi^2}{216}
 + {\rm n.p.}
\end{split}
\ee
This Ansatz ensures that $\cE^{(2)}_{(0,1)}$  satisfies the Poisson equation \eqref{eisenthree}, 
the last term in \eqref{D6R48}, proportional to $E^{SL(3)}_{[10],-3/2}\, E^{SL(2)}_{[1],3}$, being a solution to the  homogeneous equation.

At weak coupling, \eqref{D6R48} exhibits the expected perturbative contributions, up to three loops,
\be
\begin{split}
g_8^{2}\, \cE^{(2)}_{(0,1)} = & \frac{2\zeta(3)^2}{3g_8^2} + 
\left[ \frac23\zeta(3) (  \hat E^{SL(2)}_{[1],1}(T)  + \hat E^{SL(2)}_{[1],1}(U) )
+ \frac{20}{3\pi^5} E^{SL(2)}_{[1],3}(T)  E^{SL(2)}_{[1],3}(U) +\frac{\pi}{18}\zeta(3) \right]\\
+& \left[ \frac23 \hat E^{SL(2)}_{[1],1}(T)  \hat E^{SL(2)}_{[1],1}(U) 
+f(T)+f(U)+\frac{\pi}{9} ( \hat E^{SL(2)}_{[1],1}(T)  + \hat E^{SL(2)}_{[1],1}(U)  )
+\frac{11}{36}\zeta(2)
 \right] g_8^2 \\
+& \frac{2}{27} \left( E^{SL(2)}_{[1],3}(T) + E^{SL(2)}_{[1],3}(U) \right) g_8^4 
+ \frac{2\pi}{9} \left( \frac{\pi}{2}+ \cE^{(2),{\rm an}}_{(0,0)} \right) \, g_8^2 \log g_8 +  
\frac{4\pi^2}{27} g_8^2 \log^2 g_8
 + {\rm n.p.}
\end{split}
\ee
Decompactifying from $D=8$ to $D=9$, one has, in agreement 
with \eqref{decompD6R4}
\be
\begin{split}
\cE^{(2)}_{(0,1)} \rightarrow &  \left(\frac{r_2}{l_{9}}\right)^{2} \left( \cE^{(1)}_{(0,1)} 
+ \frac{15}{4\pi^3}\zeta(5)\, \left(\frac{r_2}{l_{9}}\right)^{-5} \, \cE^{(1)}_{(1,0)}
+ \frac{\pi}{36}  \left(\frac{r_2}{l_{9}}\right)^{-1}\, \cE^{(1)}_{(0,0)}
+ \tfrac{37}{36}\zeta(2)  \left(\frac{r_2}{l_{9}}\right)^{-2} \right. \\
& \left. + \frac{16\pi}{567}\zeta(6) \left(\frac{r_2}{l_{9}}\right)^5 \right) 
 + \frac{49}{27}\pi^2 (\log \left(\frac{r_2}{l_{9}}\right))^2 - \frac{7\pi}{9} \left( \left(\frac{r_2}{l_{9}}\right) \cE_{(0,0)}^{(1)}+\frac{\pi}{2} \right) \log \left(\frac{r_2}{l_{9}}\right)
\end{split}
\ee
The $\log r_2/l_9$ terms are consistent with the non-analytic term
$ \left( \frac{4\pi^2}{27} \log^2 g_8 
+(...) \log g_8 \right)$
displayed in \eqref{Enonan}.

\subsection{$D=7, d=3$}

The moduli space in type II string theory compactified on $T^3$ is $SL(5)/SO(5)$, identified under  $SL(5,\IZ)$.

\subsubsection{$\cR^4$}
The exact $\cR^4$ coupling is given by  \cite{Kiritsis:1997em}
\be
\cE^{(3)}_{(0,0)} =E^{SL(5)}_{[1000],3/2}
\ee
This reproduces the expected tree-level and one-loop terms, up to an infinite series of D-instanton corrections, 
\be
g_7^{2/5} \cE^{(3)}_{(0,0)} = \frac{2\zeta(3)}{g_7^2} +  2\pi \, E^{SO(3,3)}_{[100],1/2} + {\rm n.p.}
\ee
Note
\be
E^{SL(4)}_{[100],1}  = \pi\, E^{SO(3,3)}_{[100],1/2} 
\ee
Decompactifying from $D=7$ to $D=8$, one has, in agreement with \eqref{decompR4}
\be
\label{ER47R}
\cE^{(3)}_{(0,0)} \rightarrow  \left(\frac{r_3}{l_{8}}\right)^{6/5} \cE^{(2)}_{(0,0)} + 4\pi \left(\frac{r_3}{l_{8}}\right)^{6/5} \log \frac{r_3}{l_8 \mu_8}\
\ee
The $\log r_3$ terms cancels against the non-analytic term in $\cE^{(2)}_{(0,0)}$, so that
 $\cE^{(3)}_{(0,0)}$ is analytic at $g_7=0$. The scale is found to be $\mu_8=4\pi e^{-\gamma_E}$.

\subsubsection{$D^4 \cR^4$}
The exact $D^4 \cR^4$ coupling is given by the linear combination \cite{Green:2010wi}
\be
\cE^{(3)}_{(1,0)} =\frac12 \, \hat E^{SL(5)}_{[1000],5/2} + \frac{\pi}{30} \hat E^{SL(5)}_{[0010],5/2}
\ee
The two Eisenstein series are defined after subtracting the pole at $s=5/2$. They contribute to different supersymmetric invariants \cite{Bossard:2014aea}. At weak coupling, they reproduce
the expected tree-level, one-loop and two-loop contributions,
\be
\begin{split}
g_7^{2}\, \cE^{(3)}_{(1,0)} = & \frac{\zeta(5)}{g_7^2} +   \frac{\pi}{30} E^{SO(3,3)}_{[100],5/2}\\
&+ \left[ \frac23 \left( \hat E^{SO(3,3)}_{[010];2}+ \hat E^{SO(3,3)}_{[001];2} \right) 
+\frac{4\pi^2}{3}(1-2\gamma_E+\log 4) \right]
g_7^2  + \frac{16\pi^2}{15} 
g_7^2 \log g_7
 + {\rm n.p.}
\end{split}
\ee
Decompactifying from $D=7$ to $D=8$, one has, in agreement with \eqref{decompD4R4} \cite[4.26]{Green:2010sp},
\be
\label{ED4R47R}
\cE^{(3)}_{(1,0)} \rightarrow \left(\frac{r_3}{l_{8}}\right)^{2} \left( \cE^{(2)}_{(1,0)} 
+ \tfrac{\pi}{3 } \left(\frac{r_3}{l_{8}}\right)^{-2}
\left(  \cE^{(2)}_{(0,0)} - \tfrac{28\pi}{5} \log \frac{r_3}{l_{8}}\right)+\tfrac{\pi}{15}\zeta(5)  \left(\frac{r_3}{l_{8}}\right)^4 \right)
\ee
Using $l_8=g_8^{1/3} l_s$, one sees that the explicit $\log r_3/l_8$ term combines with the 
non-analytic contribution $\frac{4\pi}{3}\log g_8$ in $\cE^{(8)}_{(0,0)}$ to yield the non-analytic contribution $\frac{16\pi^2}{15}\, \log g_7$
in \eqref{Enonan}.

\subsubsection{$D^6 \cR^4$}

The exact $D^6\cR^4$ coupling in $D=7$ is not known. At weak coupling, it must reproduce
the correct perturbative terms up to three loops,
\be
\begin{split}
g_7^{14/5}\, \cE^{(3)}_{(0,1)} = &\frac{2\zeta(3)^2}{3g_7^2} +\left( 
\frac{2\pi\zeta(3)}{3} \, E^{SO(3,3)}_{[100],1/2} + \frac{5\pi}{378} E^{SO(3,3)}_{[100],7/2}  \right)\\
&
+  \cE_{(0,1)}^{(3,2)} g_7^2 
+ \frac{2}{27} \left[ E^{SO(3,3)}_{[001],3} +E^{SO(3,3)}_{[010],3} \right] g_7^4 
 + {\rm n.p.}
\end{split}
\ee
where $\cE_{(0,1)}^{(3,2)}$ is proportional to  the  modular integral of the Kawazumi-Zhang invariant times the genus two lattice partition function (hence not a standard Eisenstein series). 

Decompactifying from $D=7$ to $D=8$, one has, in agreement with \eqref{decompD6R4}, suitably amended to take into account the logarithmic divergences in $D=8$,
\be
\begin{split}
\cE^{(3)}_{(0,1)} \rightarrow &  \left(\frac{r_3}{l_{8}}\right)^{12/5} \left( \cE^{(2)}_{(0,1)} 
+ \frac{\pi}{18} \, \left(\frac{r_3}{l_{8}}\right)^{-4} \, \cE^{(2)}_{(1,0)}
+ \frac{2\pi}{3} \log \left(\frac{r_3}{l_{8}}\right)\, \cE^{(2)}_{(0,0)}
+ \frac{2\pi^2}{9} \log \left(\frac{r_3}{l_{8}}\right)\right.\\
& \left. +\frac{4\pi^2}{3} (\log \left(\frac{r_3}{l_{8}}\right))^2
-\frac{\pi}{36} \cE^{(2)}_{(0,0)} + \frac{5\pi^2}{72}
+ \frac{5\pi}{189}\zeta(7) \left(\frac{r_3}{l_{8}}\right)^6 \right)
\end{split}
\ee
Using $l_8=l_s g_8^{1/3}$, the $(\log \left(\frac{r_3}{l_{8}}\right))^2$ and $\log \left(\frac{r_3}{l_{8}}\right)\, \cE^{(2),an.}_{(0,0)}$ terms are seen to cancel against the non-analytic terms in $\cE^{(2)}_{(0,1)}$, so that $\cE^{(3)}_{(0,1)}$ is analytic at $g_7=0$.

\subsection{$D=6, d=4$}
The moduli space in type II string theory on $T^4$ is $SO(5,5)/SO(5)\times SO(5)$, identified under $SO(5,5,\IZ)$.

\subsubsection{$\cR^4$}
The exact $\cR^4$ coupling is given by \cite{Obers:1999um,Pioline:2010kb}
\be
\cE^{(4)}_{(0,0)} =E^{SO(5,5)}_{[10000],3/2}
\ee
At weak coupling, it produces the correct tree-level and one-loop terms,
\be
g_6 \cE^{(4)}_{(0,0)} = \frac{2\zeta(3)}{g_6^2} +  2E^{SO(4,4)}_{[1000],1} + {\rm n.p.}
\ee
Decompactifying from $D=6$ to $D=7$, one has, in agreement with \eqref{decompR4},
\be
\cE^{(4)}_{(0,0)} \rightarrow \left(\frac{r_4}{l_{7}}\right)^{3/2} \cE^{(3)}_{(0,0)} + 4\zeta(2)\, \left(\frac{r_4}{l_{7}}\right)^{5/2} 
\ee

\subsubsection{$D^4 \cR^4$}
The exact $D^4 \cR^4$ coupling is given by \cite{Green:2010wi}
\be
\cE^{(4)}_{(1,0)} =\frac12 \hat E^{SO(5,5)}_{[10000],5/2} + \frac{4}{45} \hat E^{SO(5,5)}_{[00001],3}
\ee
The two Eisenstein series are defined by subtracting the pole at $s=5/2$ and $s=3$, respectively.
They contribute to different supersymmetric invariants \cite{Bossard:2014aea}.

At weak coupling, this produces the correct tree-level, one-loop and two-loop terms,
\be
\begin{split}
g_6^{3}\, \cE^{(4)}_{(1,0)} =& \frac{\zeta(5)}{g_6^2} + 
\left( \frac{4}{45} \hat E^{SO(4,4)}_{[1000],3} + \frac{4}{3} \zeta (3) (36 \log (A)-7+3 \gamma_E )-4 \zeta '(3) \right) \\
&+ \left[ \frac23 \left( \hat E^{SO(4,4)}_{[0001],2}+\hat E^{SO(4,4)}_{[0010],2}\right) 
+\left( \frac{24 \zeta '(2)}{\pi ^2}-\frac{360 \zeta '(4)}{\pi ^4}-\frac{19}{3}+\log (16) \right)E^{SO(4,4)}_{[1000],1} 
\right] g_6^2 \\
&+ \cE^{(4)}_{(0,0)} \, g_6^{3} \log g_6
+ {\rm n.p.}
\end{split}
\ee
Decompactifying from $D=6$ to $D=7$, one has, in agreement with \eqref{decompD4R4}, suitably modified to take into account the logarithmic divergences,
\be
\label{ED4R46}
\cE^{(4)}_{(1,0)} \rightarrow \left(\frac{r_4}{l_{7}}\right)^{5/2} \left( \cE^{(3)}_{(1,0)} 
+ \pi^2\log \frac{r_4}{l_{7}}\right) -\tfrac52 \left(\frac{r_4}{l_{7}}\right)^{3/2} \log \frac{r_4}{l_7}\,  \cE^{(3)}_{(0,0)} 
+\tfrac{8}{45} \zeta(6) \left(\frac{r_4}{l_{7}}\right)^{15/2} 
\ee
Using $l_7=g_7^{2/5}l_s$, the explicit $\log r_4/l_7$ terms along with the non-analytic behavior of $ \cE^{(3)}_{(1,0)}\sim\frac{16\pi^2}{15}\log g_7$ are seen to be consistent with the non-analytic term $ \cE_{(0,0)}\, \log g_6$ in  $\cE^{(4)}_{(1,0)}$, as written  
in \eqref{Enonan}.

\subsubsection{$D^6 \cR^4$}
The exact $D^6\cR^4$ coupling in $D=6$ is not known. At weak coupling, it must reproduce
the correct perturbative terms up to three loops, 
\be
\label{D6R46}
\begin{split} 
g_6^4\, \cE^{(4)}_{(0,1)} =  &
\frac{2\zeta(3)^2}{3 g_6^{2}} 
+ \left[ \frac{2\zeta(3)}{3}  E^{SO(4,4)}_{[1000],1}
+ \frac{8}{189} E^{SO(4,4)}_{[1000],4} \right]
+ g_6^{2}  {\cE}^{(4,2)}_{(0,1)} \\
&
+ \frac{2}{27}  \left[ \hat E^{SO(4,4)}_{[0001],3}+ \hat E^{SO(4,4)}_{[0010],3}\right]g_6^4
+ 5\zeta(3) g_6^4 \log g_6 + {\rm n.p.}
\end{split}
\ee
Under decompactification from $D=6$ to $D=7$, one has, in agreement with \eqref{decompD6R4},
\be
\label{ERD6R46}
\begin{split}
\cE^{(4)}_{(0,1)} \rightarrow & 
\left(\frac{r_4}{l_7}\right)^{3}  \cE^{(3)}_{(0,1)} 
+ \frac{5\zeta(3)}{2\pi^2}  \,  \, \cE^{(3)}_{(1,0)} 
+ \frac23 \zeta(2) \left(\frac{r_4}{l_7}\right)^4\, \cE^{(3)}_{(0,0)}
+ \frac85 \zeta^2(2)  \left(\frac{r_4}{l_7}\right)^{5}  
 \\
&+ \frac{16}{189} \zeta(8) \left(\frac{r_4}{l_7}\right)^{10} 
-\frac{35}{6} \zeta(3) \log \frac{r_4}{l_7}\ .
\end{split}
\ee
Using $l_7=l_s g_7^{2/5}$, the logarithmic term is seen to combine with the non-analytic term $\frac{16\pi^2}{15}\log g_7$ in $\cE^{(3)}_{(1,0)}$ to produce $\cE^{(4)}_{(0,1)}  \sim 5\zeta(3) \log g_6$.

\subsection{$D=5, d=5$}
The moduli space in type II string theory compactified on $T^5$ is $E_{6(6)}/USp(8)$,
identified under $E_{6(6)}(\IZ)$.

\subsubsection{$\cR^4$}
The exact $\cR^4$ coupling is given by \cite{Obers:1999um,Pioline:2010kb,Green:2010kv}
\be
\cE^{(5)}_{(0,0)} =E^{E_6}_{[10^5],3/2}
\ee
At weak coupling, it produces the correct tree-level and one-loop terms,
\be
g_5^{2} \cE^{(5)}_{(0,0)} = \frac{2\zeta(3)}{g_5^2} +   E^{SO(5,5)}_{[10^4],3/2} + {\rm n.p.}
\ee
Decompactifying from $D=5$ to $D=6$, one has, in agreement with \eqref{decompR4},
\be
\cE^{(5)}_{(0,0)} \rightarrow \left(\frac{r_5}{l_{6}}\right)^{2} \cE^{(4)}_{(0,0)} + 2\zeta(3)\, \left(\frac{r_5}{l_{6}}\right)^4 
\ee

\subsubsection{$D^4 \cR^4$}
The exact $D^4 \cR^4$ coupling is given by \cite{Green:2010kv}
\be
\cE^{(5)}_{(1,0)} =\frac12 E^{E_6}_{[10^5],5/2}
\ee
At weak coupling, this produces the correct tree-level, one-loop and two-loop terms,
\be
g_5^{14/3}\, \cE^{(5)}_{(1,0)} = \frac{\zeta(5)}{g_5^2} +
\frac{1}{12} E^{SO(5,5)}_{[10^4],7/2} 
+\frac23  E^{SO(5,5)}_{[0^41],2} g_5^2 
+ {\rm n.p.}
\ee
Decompactifying from $D=5$ to $D=6$,  one has, in agreement with \eqref{decompD4R4}, suitably modified to take into account the logarithmic divergences,
\be
\cE^{(5)}_{(1,0)} \rightarrow \left(\frac{r_5}{l_{6}}\right)^{10/3} \left( \cE^{(4)}_{(1,0)} + 2\log \left(\frac{r_5}{l_{6}}\right) \, \cE^{(4)}_{(0,0)} \right)
+\tfrac{1}{6} \zeta(7)\left(\frac{r_5}{l_{6}}\right)^{28/3} 
\ee
Using $l_6=l_s g_6^{1/2}$, the $\log r_5/l_6$ term is seen to cancel against the non-analytic 
term  $ \cE_{(0,0)}\, \log g_6$ in  $\cE^{(4)}_{(1,0)}$, so that $\cE^{(5)}_{(1,0)}$ is analytic at $g_5=0$.

\subsubsection{$D^6 \cR^4$}
The exact $D^6\cR^4$ coupling in $D=5$ is not known. At weak coupling, it must reproduce
the correct perturbative terms up to three loops, 
\be
\begin{split}
g_5^{6}\, \cE^{(5)}_{(0,1)} = &\frac{2\zeta(3)^2}{3g_5^2} +
\left( \frac{\zeta(3)}{3} E^{SO(5,5)}_{[10^4],3/2} +\frac{5}{108} E^{SO(5,5)}_{[10^4],9/2} \right)
 +g_5^2 \, \cE_{(0,1)}^{(5,2)}
 \\
 &
 +\frac{2}{27} \left( \hat E^{SO(5,5)}_{[0^4 1],3} 
 +\hat E^{SO(5,5)}_{[0^310],3} \right) g_5^4 
 +\alpha_5 \,\log g_5\, \cE_{(0,0)}^{(5)}+
 {\rm n.p.}
\end{split}
\ee
Under decompactification from $D=5$ to $D=6$, one has
\be
\begin{split}
\cE^{(5)}_{(0,1)} \rightarrow &  \left(\frac{r_5}{l_{6}}\right)^{4}  \cE^{(4)}_{(0,1)} 
+ \frac56\, \left(\frac{r_5}{l_{6}}\right)^{2} \, \left( \cE^{(4)}_{(1,0)} -\frac{10}{3} \, \cE^{(4)}_{(0,0)} \log \frac{r_5}{l_{6}}\right)
+ \frac13 \zeta(3) \left(\frac{r_5}{l_{6}}\right)^6\, \cE^{(4)}_{(0,0)}\\
&+ \frac23 \zeta^2(3)  \left(\frac{r_5}{l_{6}}\right)^{8}
 + \frac{5}{54}\zeta(9) \left(\frac{r_5}{l_{6}}\right)^{12} 
+ \frac{10}{9}\zeta(3) \left(\frac{r_5}{l_{6}}\right)^{4} \log \frac{r_5}{l_{6}\tilde\mu_6}  
\end{split}
\ee

\subsection{$D=4, d=6$}

The moduli space in type II string theory is $E_{7(7)}/SU(8)$, identified under $E_{7(7)}(\IZ)$.

\subsubsection{$\cR^4$}
The exact $\cR^4$ coupling is given by \cite{Obers:1999um,Pioline:2010kb,Green:2010kv}
\be
\cE^{(6)}_{(0,0)} =E^{E_7}_{[10^6],3/2}
\ee
At weak coupling, it produces the correct tree-level and one-loop terms,
\be
g_4^{4} \cE^{(6)}_{(0,0)} = \frac{2\zeta(3)}{g_4^2} +  \frac{2}{\pi} E^{SO(6,6)}_{[10^5],2} + {\rm n.p.}
\ee
Decompactifying from $D=4$ to $D=5$, one has, in agreement with \eqref{decompR4},
\be
\cE^{(6)}_{(0,0)} \rightarrow \left(\frac{r_6}{l_{5}}\right)^{3} \cE^{(5)}_{(0,0)} + \tfrac{4}{\pi}\zeta(4)\, \left(\frac{r_6}{l_{5}}\right)^6 
\ee

\subsubsection{$D^4 \cR^4$}
The exact $D^4 \cR^4$ coupling is given by \cite{Green:2010kv}
\be
\cE^{(6)}_{(1,0)} =\frac12 E^{E_7}_{[10^6],5/2}
\ee
At weak coupling, this produces the correct tree-level, one-loop and two-loop terms,
\be
g_4^{8}\, \cE^{(6)}_{(1,0)} = \frac{\zeta(5)}{g_4^2} +
\frac{4}{15\pi} E^{SO(6,6)}_{[10^5],4} 
+\frac23  E^{SO(6,6)}_{[0^51],2}  g_4^2 
+ {\rm n.p.}
\ee
 Decompactifying from $D=4$ to $D=5$, one has, in agreement with \eqref{decompD4R4}, 
\be
\cE^{(6)}_{(1,0)} \rightarrow \left(\frac{r_6}{l_{5}}\right)^{5}  \cE^{(5)}_{(1,0)} + \tfrac{\pi}{3} \left(\frac{r_6}{l_{5}}\right)^6\, \cE^{(5)}_{(0,0)} 
+\tfrac{8}{15\pi} \zeta(8) \left(\frac{r_6}{l_{5}}\right)^{12} 
\ee

\subsubsection{$D^6 \cR^4$}
The exact $D^6\cR^4$ coupling in $D=4$ is not known. At weak coupling, it must reproduce
the correct perturbative terms up to three loops, 
\be
\begin{split}
g_4^{10}\, \cE^{(6)}_{(0,1)} = &\frac{2\zeta(3)^2}{3g_4^2} +
\left( \frac{2\zeta(3)}{3\pi} E^{SO(6,6)}_{[10^5],2} +\frac{32}{189\pi} \hat E^{SO(6,6)}_{[10^5],5} \right)
 +g_4^2 \, \cE_{(0,1)}^{(6,2)}
 \\
 &
 +\frac{2}{27} \left( \hat E^{SO(6,6)}_{[0^5 1],3} 
 +\hat E^{SO(6,6)}_{[0^410],3} \right)
 g_4^4 + \alpha_4\,  \cE^{(6)}_{(1,0)} \log g_4 +  {\rm n.p.}
\end{split}
\ee
Under decompactification from $D=4$ to $D=5$, one has
\be
\begin{split}
\cE^{(6)}_{(0,1)} \rightarrow & \left(\frac{r_6}{l_{5}}\right)^{6} \left( \cE^{(5)}_{(0,1)} + 
\frac56\,  \cE^{(5)}_{(0,0)}\, \log \left(\frac{r_6}{l_{5}}\right)
-\frac{15}{2\pi}\, \left(\frac{r_6}{l_{5}}\right)^{-1} \, \cE^{(5)}_{(1,0)}\, 
\log \left(\frac{r_6}{l_{5}}\right) \right.\\
& \left.
+ \frac{2\zeta(4)}{3\pi} \left(\frac{r_6}{l_{5}}\right)^3 \cE^{(5)}_{(0,0)}+ \frac{64\zeta(10)}{189\pi} \left(\frac{r_6}{l_{5}}\right)^{9}-\frac{4}{\pi^2}\zeta(8) \left(\frac{r_6}{l_{5}}\right)^6 \log \left(\frac{r_6}{l_{5}}\right) \right)
\end{split}
\ee

\subsection{$D=3, d=7$}
Finally, the moduli space in type II string compactified on $T^7$ is $E_{8(8)}/SO(16)$, identified under $E_{8(8)}(\IZ)$.

\subsubsection{$\cR^4$}
The exact $\cR^4$ coupling is given by \cite{Obers:1999um,Pioline:2010kb,Green:2010kv}
\be
\cE^{(7)}_{(0,0)} =E^{E_8}_{[10^7],3/2}
\ee
This reproduces the expected tree-level and one-loop terms, up to an infinite series of D-instanton corrections, 
\be
g_3^{10} \cE^{(7)}_{(0,0)} = \frac{2\zeta(3)}{g_3^2} +  \frac{3}{2\pi} E^{SO(7,7)}_{[10^6],5/2} + {\rm n.p.}
\ee
 Decompactifying from $D=3$ to $D=4$, one has, in agreement with \eqref{decompR4},
\be
\cE^{(7)}_{(0,0)} \rightarrow \left(\frac{r_7}{l_{4}}\right)^{6} \cE^{(6)}_{(0,0)} + \tfrac{3}{\pi}\zeta(5)\, \left(\frac{r_7}{l_{4}}\right)^{10} 
\ee

\subsubsection{$D^4 \cR^4$}
The exact $D^4 \cR^4$ coupling is given by \cite{Green:2010kv}
\be
\cE^{(7)}_{(1,0)} =\frac12 E^{E_8}_{[10^7],5/2}
\ee
At weak coupling, this produces the correct tree-level, one-loop and two-loop terms,
\be
g_3^{18}\, \cE^{(7)}_{(1,0)} = \frac{\zeta(5)}{g_3^2} +
\frac{7}{24\pi} E^{SO(7,7)}_{[10^6],9/2} 
+\frac23 E^{SO(7,7)}_{[0^61],2} g_3^2 
+ {\rm n.p.}
\ee
Decompactifying from $D=3$ to $D=4$,  
one has, in agreement with \eqref{decompD4R4}, 
\be
\cE^{(7)}_{(1,0)} \rightarrow \left(\frac{r_7}{l_{4}}\right)^{10}  \cE^{(4)}_{(1,0)} + \tfrac{1}{\pi}\zeta(3) \left(\frac{r_7}{l_{4}}\right)^{12}\, \cE^{(4)}_{(0,0)} 
+\tfrac{7}{12\pi} \zeta(9) \left(\frac{r_7}{l_{4}}\right)^{18} 
\ee

\subsubsection{$D^6 \cR^4$}
The exact $D^6\cR^4$ coupling in $D=3$ is not known. At weak coupling, it must reproduce
the correct perturbative terms up to three loops, 
\be
\begin{split}
g_3^{22}\, \cE^{(7)}_{(0,1)} = &\frac{2\zeta(3)^2}{3g_3^2} +
\left( \frac{\zeta(3)}{2\pi} E^{SO(7,7)}_{[10^6],5/2} +\frac{5}{24\pi} E^{SO(7,7)}_{[10^6],11/2} \right)
 +g_3^2 \, \cE_{(0,1)}^{(7,2)}
 \\
 &
 +\frac{2}{27} E^{SO(7,7)}_{[0^6 1],3} 
 g_3^4 +  {\rm n.p.}
\end{split}
\ee
Decompactifying from $D=3$ to $D=4$, one has
\be
\begin{split}
\cE^{(7)}_{(0,1)} \rightarrow & \left(\frac{r_7}{l_{4}}\right)^{12} \left( \cE^{(6)}_{(0,1)} + \frac{5}{\pi}\, \log \left(\frac{r_7}{l_{4}}\right)\,  \cE^{(6)}_{(1,0)}
+ \frac{\zeta(5)}{2\pi} \left(\frac{r_7}{l_{4}}\right)^4 \cE^{(6)}_{(0,0)} \right.\\
& \left.  + \frac{5\zeta(11)}{12\pi} \left(\frac{r_7}{l_{4}}\right)^{10} -
\frac{9\zeta(5)}{8\pi^2} \left(\frac{r_7}{l_{4}}\right)^8 \right)
\end{split}
\ee
The coefficient in front of $\log \left(\frac{r_7}{l_{4}}\right)$ is fixed by requiring that it cancels the non-analytic term $\alpha_4\, \cE_{(1,0)}^{(6)} \log g_4$ in $ \cE^{(6)}_{(0,1)}$, so that $\cE^{(7)}_{(0,1)}$ is analytic at $g_3=0$ (recall $l_4=g_4 l_s$).


\providecommand{\href}[2]{#2}\begingroup\raggedright\endgroup

\end{document}